\documentclass[epj]{svjour}
\usepackage[latin1]{inputenc}
\usepackage[dvips]{graphicx}

\usepackage{amsmath}
\usepackage{bm}
\usepackage{amssymb}
\usepackage{latexsym}
\usepackage{dcolumn}
\usepackage{url}
\newcolumntype{d}{D{.}{.}{-1}}
\newcolumntype{f}[1]{D{.}{.}{#1}}

\begin{document}
\title{QED and relativistic corrections in superheavy elements}
\author{P.\ Indelicato\inst{1}
\and
J.P.\ Santos\inst{2}
\and
S.\ Boucard\inst{1}
\and
J.-P.\ Desclaux\inst{3}
%
}                     
\offprints{P.\ Indelicato}          
\institute{
Laboratoire Kastler Brossel,
École Normale Sup\' erieure; CNRS; Université P. et M. Curie - Paris 6\\
Case 74; 4, place Jussieu, 75252 Paris CEDEX 05, France
\email{paul.indelicato@spectro.jussieu.fr}
\and
Centro de Física Atómica and  Departamento de Física, Faculdade de Ciências e Tecnologia, Universidade Nova de Lisboa,
 Monte de Caparica, 2829-516 Caparica, Portugal
\and
15 Chemin du Billery, 38360 Sassenage, France
\email{jean-paul.desclaux@wanadoo.fr}
}
\date{Received: \today / Revised version: date}
%

\abstract{In this paper we review the different relativistic and QED contributions to energies, ionic radii,
 transition probabilities and Landé $g$-factors in
super-heavy elements, with the help of the MultiConfiguration Dirac-Fock method (MCDF).
The effects of taking into account the Breit interaction to all orders by including it
in the self-consistent field process are demonstrated. State of the art radiative corrections
are included in the calculation
and discussed. We also study the non-relativistic limit of MCDF calculation and 
find that the non-relativistic offset can be unexpectedly large.
} 
\maketitle
%

\section{Introduction}
\label{intro}

In the last decades, accelerator-based experiments at GSI and Dubna have lead to the discovery
of super-heavy elements up to $Z=116$ and 118 \cite{ouya2006} (for a recent review see \cite{ham2000}).
Considerable theoretical work has been done to predict the ground configuration and the chemical properties
of those superheavy elements. Relativistic Hartree-Fock has been used to predict the ground configuration
properties of superheavy elements up to $Z=184$ in the early 70's \cite{maw1970,faw1971}. 
The Multiconfiguration Dirac-Fock (MCDF)
method was used to predict orbital properties of elements up to $Z=120$ \cite{des1973}, electron binding
energies up to $Z=118$ \cite{risp2004,srmp2006}, and K-shell and L-shell 
 ionization potentials for the superheavy elements with $Z = 112$, 114, 116, and 118 \cite{gsn2002}. Ionization
potential and radii of neutral and ionized bohrium ($Z=107$) and hassium ($Z=108$) have been evaluated
with large scale MCDF calculations \cite{jfjd2002}.
Kaldor and coworkers have employed the relativistic coupled-cluster method to predict ground state configuration, 
ionization potential, electron affinity, binding energy of the negative ion of several elements with $Z\geq 100$ 
\cite{eki1995a,eki1995b,ekip1996,sdfh1996,eki1998,leik2001a,leik2001b,elik2002}. 

Very recently, laser spectroscopy of several fermium ($Z=100$) transitions  has been performed,
the spectroscopy of nobelium ($Z=102$) is on the way \cite{sbdk2003,sbdd2003,bdfh2005}, and large scale
MCDF calculations of transition energies and rates
have been performed by several authors for superheavy elements with $Z=100$~\cite{sbdk2003,fri2005},
$Z=102$~\cite{fri2005} and $Z=103$~\cite{zaf2002}, which are in reasonable agreement with the 
fermium measurements.

There are however many unanswered questions, that need to be addressed in order to assess the accuracy and
the limit of current theoretical methods. For inner-shells, or highly ionized systems, QED effects must
 be very strong
in superheavy elements, where the atomic number $Z$ approaches the limit $Z\alpha \to 1$ ($\alpha=1/137.036$ is the
fine structure constant), at which the point-nucleus Dirac
equation eigen-energies become singular for all $s_{1/2}$ and $p_{1/2}$ states (the energy depends
 on $\sqrt{\left(j+\frac{1}{2}\right)^2-\left(Z\alpha \right)^2} $, where $j$ is the total angular momentum).
This means that all QED calculations must be performed for finite nuclei, and to all orders in $Z\alpha$.
QED calculations for outer-shell are very difficult. At present, only the simplest one-electron, one-loop diagrams
can be calculated, using model potentials to account for the presence of the other electrons.
The evaluation of many-body effects, that remains large for neutral and quasi-neutral systems, is also
made very difficult by the complex structure of these atoms, in which there may be several open shells.
In that sense, methods based on Relativistic Many-Body Perturbation theory (RMBPT) and MCDF methods
 are complementary. The former one usually allowing for more accurate results, but limited to (relativistic) closed
shell systems  minus one or plus one or at most two electrons, while the latter is completely general
but convergence becomes problematic for large size configuration set, particularly if one wants to optimize
all orbitals.

The paper is organized as follows. In Sec.\ ~\ref{sec:1} we present briefly the MCDF method, with emphasis on the
specific features of the code we have been using, and we described the QED corrections that have been used in the calculations. 
In Sec.\ ~\ref{sec:limits} we describes specific problems associated with the MCDF method (or more generally to 
all-order methods). We thus study the non-relativistic limit of the MCDF codes, and specific problems associated with high-$Z$.
In Sec.\ ~\ref{sec:eff-breit} we study a number of systems from highly-charges ions to neutral atoms, for very large atomic numbers.
The evaluation of atomic charge distribution size and Landé factors is performed in Sec.\ ~\ref{sec:land-rad} and  in Sec.\ ~\ref{sec:concl} we state
 our conclusion.

\section{Calculation of atomic wavefunctions and transition probabilities}
\label{sec:1}
\subsection{The MCDF method}
\label{sec:mcdf}

In this work, bound-states wavefunctions are calculated using the 2006 version of the Dirac-Fock program of J.-P.\ ~Desclaux and P.\ ~Indelicato,
 named \emph{mdfgme}~\cite{iad2005}. Details on the Hamiltonian and the processes
used to build the wave-functions can be found elsewhere~\cite{des1975,des1993,ind1995,ind1996}.

The total wavefunction is calculated with the help of the variational
principle. The total energy of the atomic system is the eigenvalue of the
equation
\begin{equation}
\label{eq:hamil}
        {\cal H}^{\mbox{{\tiny no pair}}}
        \Psi_{\Pi,J,M}(\ldots,\bm{r}_{i},\ldots)=E_{\Pi,J,M}
        \Psi_{\Pi,J,M}(\ldots,\bm{r}_{i},\ldots),
\end{equation}
where $\Pi$ is the parity, $J$ is the total angular momentum eigenvalue, and
$M$ is the eigenvalue of its projection on the $z$ axis $J_{z}$. 
Here,
\begin{equation}
\label{eq:hamilnopai}
        {\cal H}^{\mbox{{\tiny no pair}}}=\sum_{i=1}^{N}{\cal
        H}_{D}(r_{i})+\sum_{i<j}{\cal V}(|\bm{r}_{i}-\bm{r}_{j}|),
\end{equation}
where ${\cal H}_{D}$ is the one electron Dirac operator and ${\cal V}$
is an operator representing the electron-electron interaction of order
one in $\alpha$. 
The expression of $V_{ij}$ in Coulomb gauge, and in atomic units,is
%
\begin{subequations}
\label{eq:eeinter}
\begin{align}
         V_{ij} =& \,\,\,\, \frac{1}{r_{ij}} \label{eq:coulop} \\
         &-\frac{\bm{\alpha}_{i} \cdot \bm{\alpha}_{j}}{r_{ij}} 
\label{eq:magop} \\ 
         & - \frac{\bm{\alpha}_{i} \cdot
         \bm{\alpha}_{j}}{r_{ij}} 
[\cos\left(\frac{\omega_{ij}r_{ij}}{c}\right)-1]
         \nonumber \\
        & + c^2(\bm{\alpha}_{i} \cdot
         \bm{\nabla}_{i}) (\bm{\alpha}_{j} \cdot
         \bm{\nabla}_{j})
         \frac{\cos\left(\frac{\omega_{ij}r_{ij}}{c}\right)-1}{\omega_{ij}^{2} 
r_{ij}},
         \label{eq:allbreit}
\end{align}
\end{subequations}
%
where $r_{ij}=\left|\bm{r}_{i}-\bm{r}_{j}\right|$ is the
inter-electronic distance, $\omega_{ij}$ is the energy of the
exchanged photon between the two electrons, $\bm{\alpha}_{i}$ are the
Dirac matrices and $c$ is the speed of light. We use the Coulomb gauge
as it has been demonstrated that it provides energies free from
spurious contributions at the ladder approximation level and must be
used in many-body atomic structure calculations~\cite{gai1988,lam1989}.

The term (\ref{eq:coulop}) represents the Coulomb interaction, the
term (\ref{eq:magop}) is the Gaunt (magnetic) interaction, and the
last two terms (\ref{eq:allbreit}) stand for the retardation operator.
In this expression the $\bm{\nabla}$ operators act only on $r_{ij}$
and not on the following wavefunctions.

By a series expansion of the operators in expressions~(\ref{eq:magop})
and (\ref{eq:allbreit}) in powers of $\omega_{ij}r_{ij}/c \ll 1$ one
obtains the Breit interaction, which includes the leading retardation
contribution of order $1/c^{2}$. The Breit interaction is, then, the
sum of the Gaunt interaction (\ref{eq:magop}) and the Breit
retardation
\begin{equation}
\label{eq:breit}
B^{\text{\scriptsize{R}}}_{ij} =
{\frac{\bm{\alpha}_i\cdot\bm{\alpha}_j}{2r_{ij}}} - 
\frac{\left(\bm{\alpha}_i\cdot\bm{r}_{ij}\right)\left(\bm{\alpha}_j
\cdot\bm{r}_{ij}\right)}{{2r_{ij}^3}}.
\end{equation}
In the many-body part of the calculation the electron-electron
interaction is described by the sum of the Coulomb and the Breit
interactions. Higher orders in $1/c$, deriving from the difference
between expressions (\ref{eq:allbreit}) and (\ref{eq:breit}) are treated here
only as a first order perturbation.
All calculations are done for finite nuclei using a Fermi distribution with a
tickness parameter of 2.3 fm. The nuclear radii are taken or evaluated using formulas
from reference~\cite{ang2004}.

The MCDF
method is defined by the particular choice of a trial function to solve equation
(\ref{eq:hamil}) as a linear combination of configuration state functions (CSF):
\begin{equation}
\left\vert \mathit{\Psi}_{\mathit{\Pi},J,M}\right\rangle =\sum_{\nu=1}%
^{n}c_{\nu}\left\vert \nu,\mathit{\Pi},J,M\right\rangle . \label{eq_cu}%
\end{equation}
The CSF are also eigenfunctions of the parity $\mathit{\Pi}$, the total
angular momentum $J^{2}$ and its projection $J_{z}$. The label $\nu$ stands
for all other numbers (principal quantum number, ...) necessary to define
unambiguously the CSF. The $c_{\nu}$ are called the mixing coefficients and
are obtained by diagonalization of the Hamiltonian matrix coming from the
minimization of the energy in equation~(\ref{eq:hamil}) with respect to the $c_{\nu}%
$.
The CSF are antisymmetric products of one-electron wavefunctions
expressed as linear combination of Slater determinants of Dirac
4-spinors

\begin{equation}
\left\vert \nu,\mathit{\Pi},J,M\right\rangle =\sum_{i=1}^{N_{\nu}}%
d_{i}\left\vert
\begin{array}
[c]{ccc}%
\psi_{1}^{i}\left(  \boldsymbol{r}_{1}\right)  & \cdots & \psi_{m}^{i}\left(\boldsymbol{r}_{1}\right)
\\
\vdots & \ddots & \vdots\\
\psi_{1}^{i}\left(  \boldsymbol{r}_{m}\right)  & \cdots & \psi_{m}^{i}\left(  \boldsymbol{r}_{m}\right)
\end{array}
\right\vert ,
\end{equation}
where the $\psi$-s are the one-electron wavefunctions and the coefficients
$d_{i}$ are determined by requiring that the CSF is an eigenstate of $J^{2}$
and $J_{z}$.
The one-electron wavefunctions are defined as
\begin{equation}
\psi\left(r\right) = \left(
\begin{array}
[c]{c}%
\chi^{\mu}_{\kappa}(\Omega) P(r) \\
i \chi^{\mu}_{-\kappa}(\Omega) Q(r)
\end{array}
\right),
\label{eq:dir-wf}
\end{equation}
where $\chi^{\mu}_{\kappa}$ is a two-component spinor, and $P$ and $Q$ are respectively the large and small component of the wavefunction.

Application of the  variational principle leads to a set of integro-differential equations, which
determines the radial wavefunctions and a Hamiltonian matrix, which provides the
mixing coefficients $c_{\nu}$ by diagonalization.
The \emph{mdfgme} code provide the possibility to obtain wavefunctions and mixing coefficient
with either only the Coulomb \eqref{eq:coulop} interaction used to obtain the differential equations and the Hamiltonian
matrix that is diagonalized to obtain mixing coefficient or the full Breit operator \eqref{eq:breit}. 
The convergence process is based on the self-consistent field process (SCF). 
For a given set of configurations, initial wavefunctions, obtained for example, with a Thomas-Fermi potential,
are used to derive the Hamiltonian matrix and set of mixing coefficients. Direct and exchange
potential are constructed for all orbitals, and the differential equations are solved.
Then a new set of potentials is constructed and the whole process is repeated.
Each time the largest variation of all wavefunction has been reduced by an order of magnitude, 
a new Hamiltonian matrix is build and diagonalized, and a new cycle is done.

The so-called Optimized Levels (OL) method was used to determine the wavefunction
 and energy for each state involved. This allow for a full relaxation of the initial and final states
and provide much better energies and wavefunctions. However, in this method, spin-orbitals in the
initial and final states are not orthogonal, since they have been optimized
separately. The formalism to take in account the wavefunctions
non-orthogonality in the transition probabilities calculation has been
described by L\"{o}wdin \cite{low1955a} and Slater \cite{sla1963}. The matrix element of a one-electron
operator $O$ between two determinants belonging to the initial and final
states can be written as
\begin{align}
&  \left\langle \nu\mathit{\Pi}JM\right\vert \sum_{i=1}^{N}O\left(
r_{i}\right)  \left\vert \nu^{\prime}\mathit{\Pi}^{\prime}J^{\prime}M^{\prime
}\right\rangle = \nonumber\\
& ×  \frac{1}{N!}   \left\vert
\begin{array}
[c]{ccc}%
\psi_{1}\left(  r_{1}\right)  & \cdots & \psi_{m}\left(  r_{1}\right) \\
\vdots & \ddots & \vdots\\
\psi_{1}\left(  r_{m}\right)  & \cdots & \psi_{m}\left(  r_{m}\right)
\end{array}
\right\vert \nonumber \\
& × \sum_{i=1}^{m}O\left(  r_{i}\right)  \left\vert
\begin{array}
[c]{ccc}%
\phi_{1}\left(  r_{1}\right)  & \cdots & \phi_{m}\left(  r_{1}\right) \\
\vdots & \ddots & \vdots\\
\phi_{1}\left(  r_{m}\right)  & \cdots & \phi_{m}\left(  r_{m}\right)
\end{array}
\right\vert , \label{eq002}
\end{align}
where the $\psi_{i}$ belong to the initial state and the $\phi_{i}$ and primes
belong to the final state. If $\psi=\left\vert n\kappa\mu\right\rangle $ and
$\phi=\left\vert n^{\prime}\kappa^{\prime}\mu^{\prime}\right\rangle $ are
orthogonal, i.e., $\left\langle n\kappa\mu|n^{\prime}\kappa^{\prime}%
\mu^{\prime}\right\rangle =\delta_{n,n^{\prime}}\delta_{\kappa,\kappa^{\prime
}}\delta_{\mu,\mu^{\prime}}$, the matrix element (\ref{eq002}) reduces to one
term $\left\langle \psi_{i}\right\vert O\left\vert \phi_{i}\right\rangle $
where $i$ represents the only electron that does not have the same
spin-orbital in the initial and final determinants. Since $O$ is a
one-electron operator, only one spin-orbital can change, otherwise the matrix
element is zero. In contrast, when the orthogonality between initial and final
states is not enforced, one gets~\cite{low1955a,sla1963}
\begin{equation}
\left\langle \nu\mathit{\Pi}JM\right\vert \sum_{i=1}^{N}O\left(  r_{i}\right)
\left\vert \nu^{\prime}\mathit{\Pi}^{\prime}J^{\prime}M^{\prime}\right\rangle
=\sum_{i,j^{\prime}}\left\langle \psi_{i}\right\vert O\left\vert
\phi_{j^{\prime}}\right\rangle \xi_{ij^{'}} D_{ij^{\prime}},
\end{equation}
where $D_{ij^{\prime}}$ is the minor determinant obtained by crossing out the
$i$th row and $j^{\prime}$th column from the determinant of dimension $N×
N$, made of all possible overlaps $\left\langle \psi_{k}|\phi_{l^{\prime}%
}\right\rangle $ and $\xi_{ij^{'}}=± 1$ the associated phase factor. 

The \emph{mdfgme} code take into account non-orthogonality for all one-particle off-diagonal operators (hyperfine
matrix elements, transition rates\ldots). The overlap matrix is build and stored, and minor determinants are constructed,
and calculated using standard LU decomposition.

\subsection{Evaluation of QED corrections}
\label{sec:qed}

In superheavy elements, the influence of radiative corrections must be carefully studied. Obviously 
the status of the inner orbital and of the outer ones is very different. It is not possible
for the time being, to do a full QED treatment. Here we use the
one-electron self-energy  obtained using the method developed by Mohr
\cite{moh1974a,moh1974b}. These calculations have been extended first to the $n=2$ shell \cite{moh1975,moh1982} 
and then to the $n=3$, $n=4$ and  $n=5$ shells, for $|\kappa|\leq 2$ \cite{mak1992}. More recently, a new coordinate-space 
renormalization 
method has been developed by Indelicato and Mohr, that has allowed substantial gains in accuracy and
 ease of extension \cite{iam1992,iam1998}. Of particular interest for the present work, is the extension
of these calculation to arbitrary $\kappa$ values and large principal quantum numbers \cite{lim2001}. All
known values to date have been implemented in the 2006 version of the \emph{mdfgme} code, including less accurate, inner shell ones,
that covers the superheavy elements \cite{caj1976,ssmg1992}. The self-energy  of the $1s$, $2s$ and $2p_{1/2}$ states is
 corrected for finite nuclear size~\cite{mas1993}. 
The self-energy screening  is taken into account here by the Welton method \cite{igd1987,iad1990},
 which reproduces very well
other methods based on direct QED evaluation of the one-electron self-energy diagram with
 screened potentials \cite{blu1992,blu1993a,blu1993b}. Both
methods however leave out reducible and vertex contributions. These two contributions, however,
cancels out in the direct evaluation of the complete set of one-loop screened self-energy diagram with one photon exchange \cite{iam2001}.
The advantages of screening method, on the other hand is that they go beyond one photon exchange, which may be
important for the outer shells of neutral atoms. Recently special studies of outer-shell 
screening have been performed for alkali-like elements, using the multiple
 commutators method  \cite{ptl1998,lgtp1999}.

The comparison between the Welton model and the results from Ref.\ ~\cite{ptl1998,lgtp1999} is presented in Table
\ref{tab:sescreen}. This table confirms comparison with earlier work at lower $Z$. It shows
 that the use of a simple scaling law, as incorporated in GRASP 92 and earlier version of \emph{mdfgme} does not provide correct
values. This scaling  law is obtained by comparing the mean value of the radial coordinate over Dirac-Fock radial wave-function $\langle r\rangle_{\mathrm{DF}}$
to the  hydrogenic one $\langle r\left(Z_{\mathrm{eff}}\right)\rangle_{\mathrm{hydr.}}$. This allow to derive an effective atomic number $Z_{\mathrm{eff}}$  by solving
 $\langle r\left(Z_{\mathrm{eff}}\right)\rangle_{\mathrm{hydr.}}=\langle r\rangle_{\mathrm{DF}}$.  One then use $Z_{\mathrm{eff}}$ to evaluate the self-energy screening from one-electron self-energy calculations.
The superiority of the Welton model can be easily explained by noticing that the range of QED corrections is the electron Compton wavelength $\Lambda_{C}=\alpha$~a.u., while 
mean atomic orbital radii are dominated by contributions from the  $\frac{n^2}{Z}$~a.u. range, which is much larger.

\begin{table}
\begin{center}
\caption{Self-energy and self-energy screening for element 111\label{tab:sescreen} }
\begin{minipage}{\columnwidth}
\begin{tabular}{ldd}
\hline
\hline
Level  &  \multicolumn{1}{c}{$1s$}  &  \multicolumn{1}{c}{$7s$}  \\
\hline
SE (point nucl.)  &  848.23  &  3.627  \\
Welton screening  &  -18.80  &  -3.283  \\
Finite size  &  -50.37  &  -0.260  \\
Total SE (DF)  &  779.06  &  0.084  \\
Pyykkö et al. \cite{lgtp1999}\footnote{screening calculated using Dirac-Fock potential}  &    &  0.087  \\
Pyykkö et al. \cite{lgtp1999}\footnote{screening calculated using Dirac-Slater potential fitted to $E_{DF}$}  &    &  0.095  \\
$\langle r\rangle $ (GRASP) \cite{pfg1996}\footnote{use hydrogenic values with $Z_{\mathrm{eff}}$ obtained by solving
 $\langle r\left(Z_{\mathrm{eff}}\right)\rangle_{\mathrm{hydr.}}=\langle r\left(Z_{\mathrm{eff}}\right)\rangle_{\mathrm{DF}}$} &    &  0.018  \\
\hline       
\hline       
\end{tabular}       
\end{minipage}
\end{center}       
\end{table}       
%

When dealing with very heavy elements in the limit $Z\alpha \to 1$, one should consider if perturbative QED
is still a valid model. Unfortunately, we still lack the tools to answer this fundamental question.
However, the use of \emph{numerical all-order methods} may gives some partial answers. In particular they
allow to include the leading contribution to vacuum polarization to all orders by adding the
Uehling potential \cite{ueh1935} to the MCDF differential equations. This possibility has been implemented in
the \emph{mdfgme} code as described in \cite{bai2000}, with the help of Ref.\ ~\cite{kla1977}. It also allows to calculate the effect of vacuum polarization 
in quantities other than energies, like hyperfine structure shifts \cite{bai2000}, Landé $g$-factors \cite{imqd2003}, or transition rates. It can also provides some hints on
the oder of magnitude of QED effects on atomic wavefunction, orbital or atom radii, and electronic densities.

For high-$Z$, higher order QED corrections are also important. In the last decade, calculations have provided the complete
set of values for two-loops, one-electron diagrams to all orders in $Z\alpha$ \cite{yis2003a,yis2003b,yis2005a,yis2005b,yis2006}
and also low-$Z$ expansions. All available data has been implemented in the mdfgme code. However, this data is 
limited to the $n\leq 2$ shells.

\section{Limitation of the MCDF method}
\label{sec:limits}

\subsection{Non-relativistic limit}
\label{sec:nonrel}
The success of relativistic calculations in high-$Z$ elements atomic structure is impressive. It has been shown many times,
that only a fully relativistic formalism and the use of a \emph{fully} relativistic electron-electron interaction
 can reproduce the correct level ordering and energy in heavy systems. As a non-exhaustive list of example, we can
cite the case of the $1s2p\;^3P_0$--$^3P_1$ level crossing for $Z=47$ in heliumlike systems \cite{jal1976,jas1986,ipm1989},
 and the prediction of the $1s^22s 2p\;^3P_0$--$^3P_1$ inversion in Be-like iron \cite{dck1979}, both due to relativistic effects and
Breit interaction.
Relativistic effects determine also the structure of the ground configuration of many systems, as was recognized for example
in the study of lawrencium which has a $7p\;^2P$ in place of a $6d\;^2D$ configuration\cite{van1972,nsfm1974,daf1980}.

There is one caveat that must be taken into consideration when performing such calculations, that has been recognized in
low-$Z$ systems, but never explored in the super-heavy elements region:
it may sound rather paradoxical to investigate the nonrelativistic limit of MCDF and, more generally, of all-order
calculations, when studying superheavy elements. 
Here we show, however, that there is a problem that has to be taken into account  if one wants to obtain the correct
fine structure splitting in all cases. We believe it is the first time this problem is recognized in the highly-relativistic limit.

This problem was first found many years ago, in systems like the fluorine
isoelectronic sequence \cite{hkcd1982}: the non-relativistic limit, obtained by doing $c\to \infty$ in a MCDF code,
is not properly recovered. States with LSJ label $^{2S+1}L_J$ and identical $L$ and $S$, and different $J$, which
 should have had the same energy in the non-relativistic limit do not. The energy difference between the levels of identical LS labels but different $J$
is called the non-relativistic (NR) offset. This offset leads to slightly incorrect fine structure in cases when 
a relativistic configuration has several non-relativistic parents (i.e., several
states with one electron less, and different angular structure, that can recouple to give the same configuration).
 This effect can be large enough to affect comparison between theory and experiment. It should be noted that such a
 problem does not show if one works in the Extended Average Level 
(EAL) version of the MCDF. In this case a single wavefunction is used for all the members of a given multiplet, and the relaxation effects that are the source of 
the NR offset disappear, at the price of less accurate transition rates and energies, for a given configuration space.

\begin{figure}[htbp]
\centering
\includegraphics[width=\columnwidth]{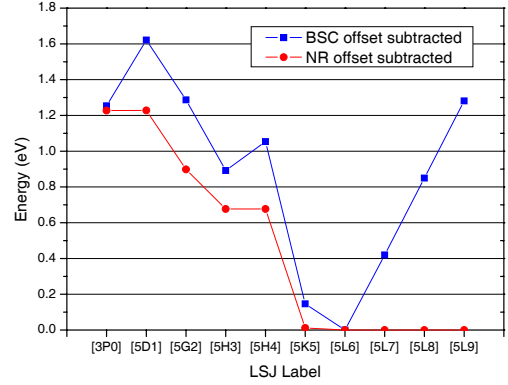}
\caption{Non-relativistic offset on the ground configuration of uranium.
``Ener. (BSC off. sub)'': total MCDF energy, with
all QED corrections, using the full Breit operator in the SCF process, relative to the $5f^3 6d 7s^2\;^5L_6$ energy to which the non-relativistic
offset has been subtracted.
 ``Ener. (NR off. sub)'': Total non-relativistic energy, to which the non-relativistic offset for members of the same 
LS multiplet has been subtracted, relative to the $5f^3 6d 7s^2\;^5L_6$ energy. The two $^5H_J$ and the four $^5L_J$ levels have thus identical
non-relativistic energy as it should be. Uncorrected energy and NR offset are displayed in \protect Table~\ref{tab:uran-nr-off} }
\label{fig:uranium-cinf}
\end{figure}

Very recently, this non-relativistic problem was shown to be general to any all-order
methods, in which sub-classes of many-body diagrams are re-summed, without including all the diagrams relevant of a given
order. In the MCDF method, in particular, one should add all configurations with single excitations of the
kind $n\kappa \to n'\kappa$, that in principle, should have no effect on the energy in a non-relativistic calculation, due
to Brillouin's theorem \cite{ild2005}. In the iso-electronic sequence investigated up to now, this effects became 
less severe when going to higher $Z$, and it has thus never been considered in very heavy systems:
 in neutral, or quasi-neutral superheavy elements, the outer-shell structure can
be complex, with several open shells, and thus many possible parents core configurations.
 It is then worthwhile to study if this problem could arise. We have  studied a number of cases. As a first example
we have studied neutral uranium. The ground state configuration is known to be [Rn]­$5f^3 6d 7s^2\;^5L_6$. We have calculated
all levels of the ground configuration with $J=0$ to $J=9$, both in normal conditions, and taking the speed of light to
infinity in the code. The results are shown on Table~\ref{tab:uran-nr-off} and Fig.\ ~\ref{fig:uranium-cinf}. There are two group of levels that can be
affected by a NR offset: the $5f^3 6d 7s^2\;^5H_J$, $J=3$, 4 and the  $5f^3 6d 7s^2\;^5L_J$, $J=6$ to $J=9$.
The non-relativistic offset is evaluated for both groups of levels 
as the difference between the energy of a  configuration with a given $J$ and the one for the configuration
with the lower energy in the NR limit. The figure clearly shows a  NR offset of 25~meV for the $^5H_4$ and $^5H_5$, and 
one of less than 1~meV for the four $^5L_J$ levels. While it can be significant compared to the accuracy of a laser measurement,
it is probably negligible compared to the accuracy of realistic correlation calculations.

\begin{figure}[htbp]
\centering
\includegraphics[width=\columnwidth]{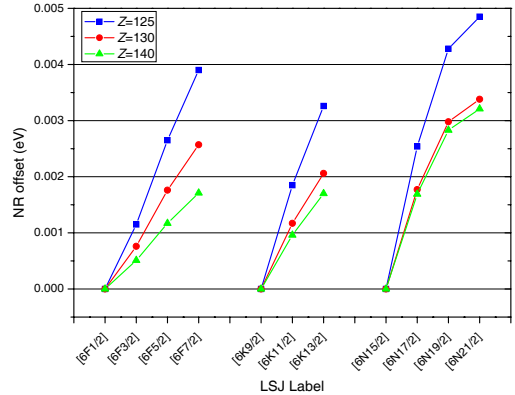}
\caption{Non-relativistic offset for the $^6F_J$, $^6K_J$ and $^6N_J$ LS configurations for the ground configuration of an atom with 125 electron. 
$Z=125$, 130 and 1340 have been evaluated. This contribution should be negligible compared to correlation.
For each LS group, the offset is evaluated by subtracting the lower energy of the group to the others.
 }
\label{fig:elem125-nro}
\end{figure}

In order to assess the generality of this problem, we have investigated several other characteristic systems.
Element 125, for example,  is the first element with a populated $5g$ orbital \cite{maw1970,faw1971}. We have calculated the NR offset
for a configuration with 125 electrons [Rn]­$5f^{14} 6d^{10}    7s^{2}  7p^6 8s^2  5g 6f^4$ and $Z=125$, 135 and 140.
The results are presented on Fig.\ ~\ref{fig:elem125-nro}. We find three groups of levels with LSJ labels, four
levels with label $^6F_J$ ($J=1/2$ to $J=7/2$),  three levels with label $^6K_J$ ($J=9/2$ to $J=13/2$) and four levels with label
$^6N_J$ (J=$15/2$ to $J=21/2$). The NR offset in each group is of the order of a few meV, while the (non-relativistic) energy
difference between the two first groups is 0.19~eV, and between the first and the last groups is around 0.25~eV. 
The figure also shows that the NR offset gets smaller when $Z$ increases as expected.

\begin{table}[htbp]
\begin{center}
\caption{Total relativistic energy [Ener. (BSC)], including all-order Breit and
QED corrections, total non-relativistic energy [ Ener. (NR)], and NR Offset for the ground 
configuration of uranium, relative to the $5f^3 6d 7s^2\;^5L_6$ energy, which is the lowest level 
of the configuration. Correlation has not been included. One can observe a splitting in the NR energy between the $^5H$ levels
and the $^5L$ levels, which should be exactly degenerate. The
 fine-structure can be improved  by subtract the NR offset from the relativistic energy.
\label{tab:uran-nr-off} }
\begin{tabular}{crrr}
\hline       
\hline       
Label & \multicolumn{1}{c}{Ener. (BSC)} & Ener. (NR) & NR Offset \\
\hline       
$^3P_0$ & 1.25373 & 1.22751 &  \\
$^5D_1$ & 1.62153 & 1.22803 &  \\
$^5G_2$ & 1.28675 & 0.89836 &  \\
\hline       
$^5H_3$ & 0.91680 & 0.70185 & 0.0251 \\
$^5H_4$ & 1.05358 & 0.67678 & 0.0000 \\
\hline       
$^5K_5$ & 0.14628 & 0.01169 &  \\
\hline       
$^5L_6$ & 0.00000 & 0.00000 & 0.0000 \\
$^5L_7$ & 0.42167 & 0.00181 & 0.0018 \\
$^5L_8$ & 0.85283 & 0.00333 & 0.0033 \\
$^5L_9$ & 1.28393 & 0.00284 & 0.0028 \\
\hline       
\hline       
\end{tabular}       
\end{center}       
\end{table}       


We have also investigated the lower excited states of a somewhat simpler system, element 118 (eka-radon). 
We have explored the [Rn]­$ 5f^{14} 6d^{10} 7s^2 7p^5 8s$ which should exhibit no NR offset (it has single parent states)
and [Rn]­$ 5f^{14} 6d^{10} 7s^2 7p^5 7d$. The results are presented on Table  ~\ref{tab:elem118-nro}. As expected,
the $7p^5 8s\;^3P_J$ states do not exhibit an NR offset, within our numerical accuracy. The  $7p^5 7d\;^3L_J$ configurations
however do have a strong NR offset, up to 0.8~eV, much larger than we expected from the other results presented above,
and the largest ever observed.
Clearly, such a large offset would render any calculation of
the fine structure splitting of eka-radon useless, unless the results are corrected for the NR offset.



\begin{table}
\begin{center}
\caption{Non-relativistic offset on the lower excited configurations of eka-radon (Z=118).
For each member of a multiplet, as, e.g., $7p^5 7d \;^3D$, we evaluate the difference 
between the member with the lowest non-relativistic energy and the others. Non-relativistically, all
member of a multiplet with identical LS labels should have an energy independent of the total angular momentum 
$J$. 
\label{tab:elem118-nro} }
\scriptsize{
\begin{tabular}{crcrcrcr}
\hline
\hline
$J$ & $7p^5 7d \;^3D$ & $J$ & $7p^5 7d \;^3F$ & $J$ & $7p^5 7d\;^ 3P$ & $J$ & $7p^5 8s \;^3P$ \\
\hline
1 & 0.801 & 2 & 0.335 & 0 & 0.000 & 0 & 0.000 \\
2 & 0.000 & 3 & 0.000 & 1 & 0.002 & 1 & 0.002 \\
3 & 0.336 & 4 & 0.000 & 2 & 0.802 & 2 & 0.002 \\
\hline       
\hline       
\end{tabular} }      
\end{center}       
\end{table}       

We would like to note that subtracting the NR offset is only a partial fix, since it was shown in Ref.\ ~\cite{ild2005} that not only the
fine structure is affected, but also the level energy. The only possible solution are thus to do calculations
with a large number of configurations, including all single excitations, or to use the EAL method.
In view of the complexity of calculations on the superheavy elements, and of the extra convergence difficulties
 associated with the presence 
of Brillouin excitations, the first solution is probably not very universal, but the second one should always work.
In conclusion, we have shown for the first time, that it is important to check for the non-relativistic limit in 
super-heavy elements and to correct for the non-relativistic offset, as it can be large in some cases.

\subsection{Effect of the all-order Breit operator on simple systems}
\label{sec:eff-breit}
\begin{figure}[htbp]
\centering
\includegraphics[width=\columnwidth]{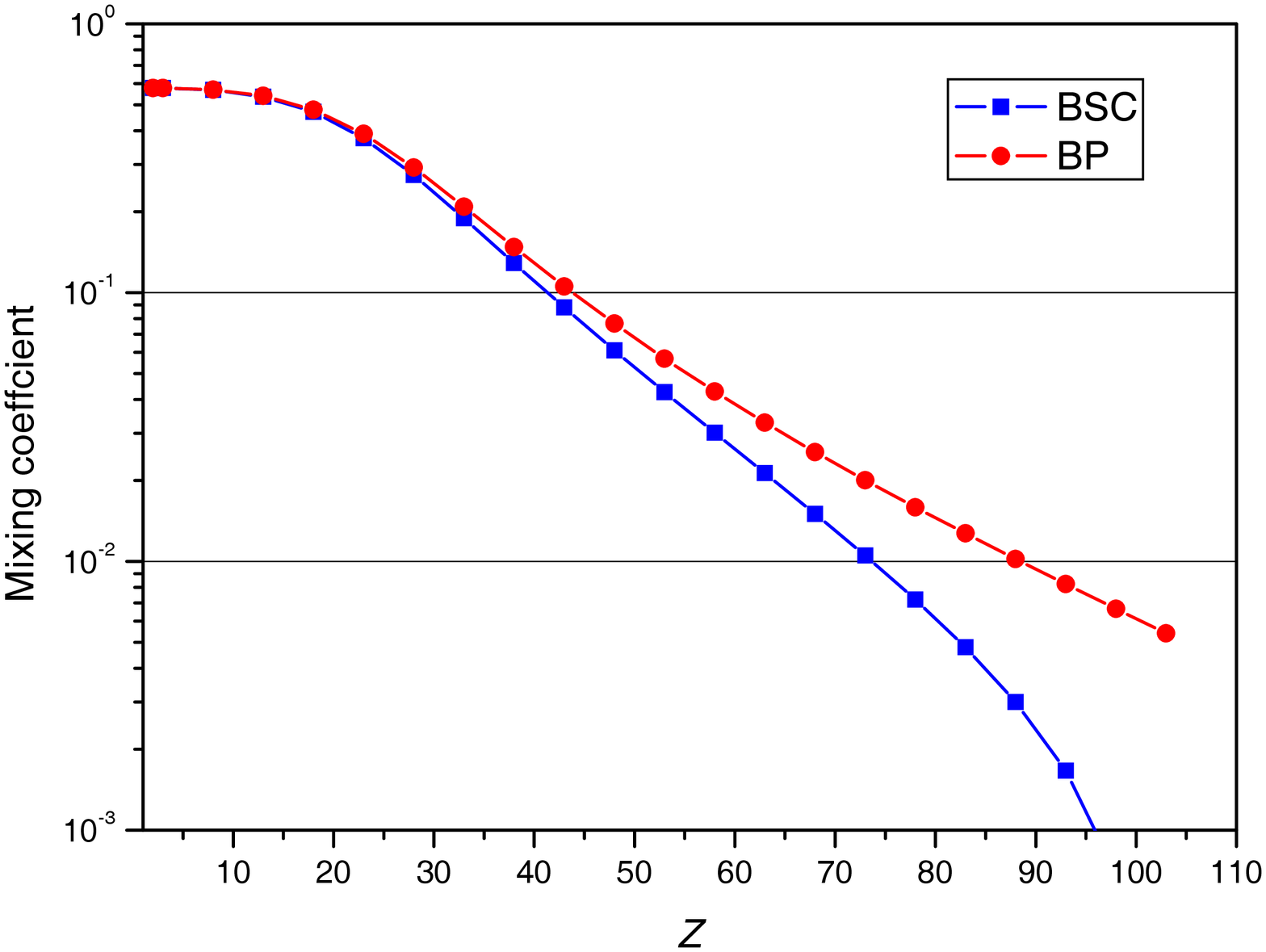}
\caption{Variation with $Z$ of mixing $c_2$ coefficient for the $1s 2p\;^3P_1$ level of helium-like ions. BP: Only the Coulomb interaction
is used in the SCF process. BSC: The full Breit interaction is used in the SCF process}
\label{fig:jj-coupling}
\end{figure}

\begin{figure*}[htbp]
\centering
\includegraphics[width=\textwidth]{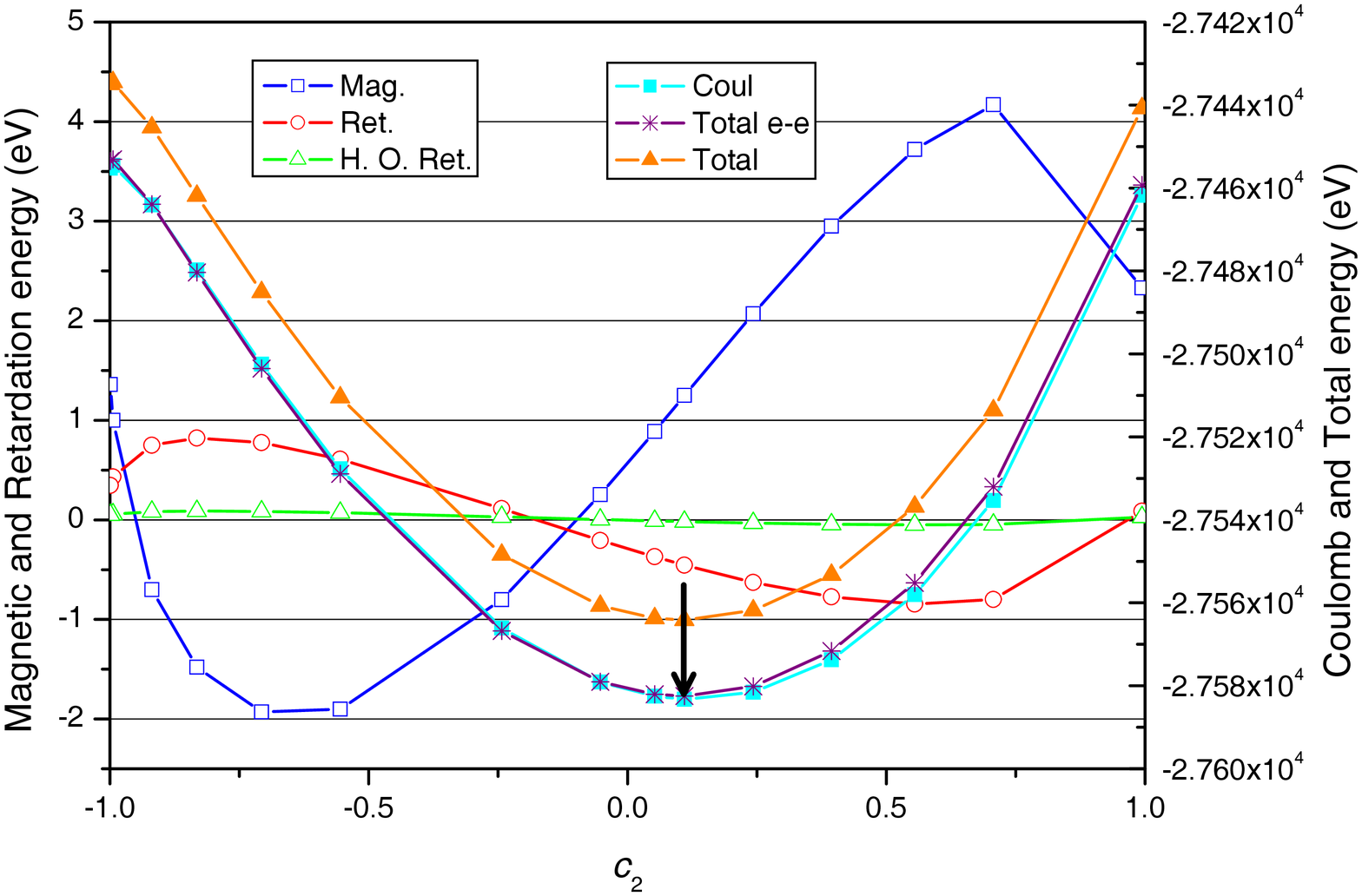}
\caption{Variation of the different contributions to the energy as a function of the mixing coefficient $c_2$, for
the $1s 2p\;^3P_1$ level at $Z=40$. The arrow indicates the position of the $1s 2p \;^3P_1$ energy, at the minimum
 around $c_2=0.104$. \textbf{Left axis:} ``Mag.'': Magnetic energy, Eq. \eqref{eq:magop}. ``Ret.'': Breit retardation,
 Eq. \eqref{eq:breit}. ``H.O. Ret.'': higher-order retardation. \textbf{Right axis:} ``Coul.'': Coulomb
energy Eq. \eqref{eq:coulop}. ``Total e-e'': sum of the 4 preceding contributions.
`` Total'': total level energy including all QED corrections. }
\label{fig:z40-3p1}
\end{figure*}

\begin{figure*}[htbp]
\centering
\includegraphics[width=\textwidth]{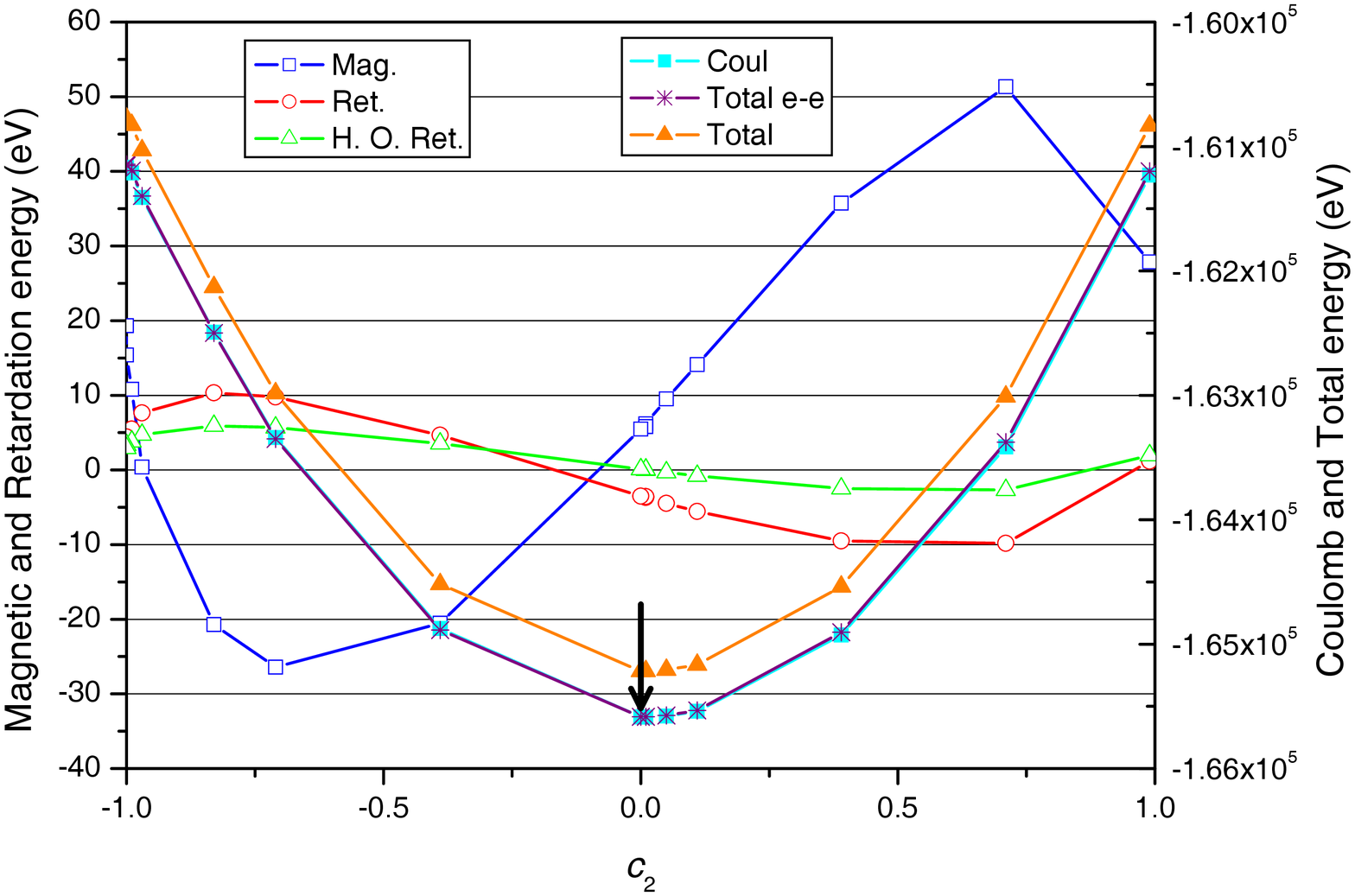}
\caption{Variation of the different contributions to the energy as a function of the mixing coefficient $c_2$, for
the $1s 2p\;^3P_1$ level at $Z=92$. The arrow indicates the position of the $1s 2p \;^3P_1$ energy, at the minimum
 around $c_2=0.002$. \textbf{Left axis:} ``Mag.'': Magnetic energy, Eq. \eqref{eq:magop}. ``Ret.'': Breit retardation,
 Eq. \eqref{eq:breit}. ``H.O. Ret.'': higher-order retardation. \textbf{Right axis:} ``Coul.'': Coulomb
energy Eq. \eqref{eq:coulop}. ``Total e-e'': sum of the 4 preceding contributions.
`` Total'': total level energy including all QED corrections.}
\label{fig:z92-3p1}
\end{figure*}

\begin{figure*}[htbp]
\centering
\includegraphics[width=\textwidth]{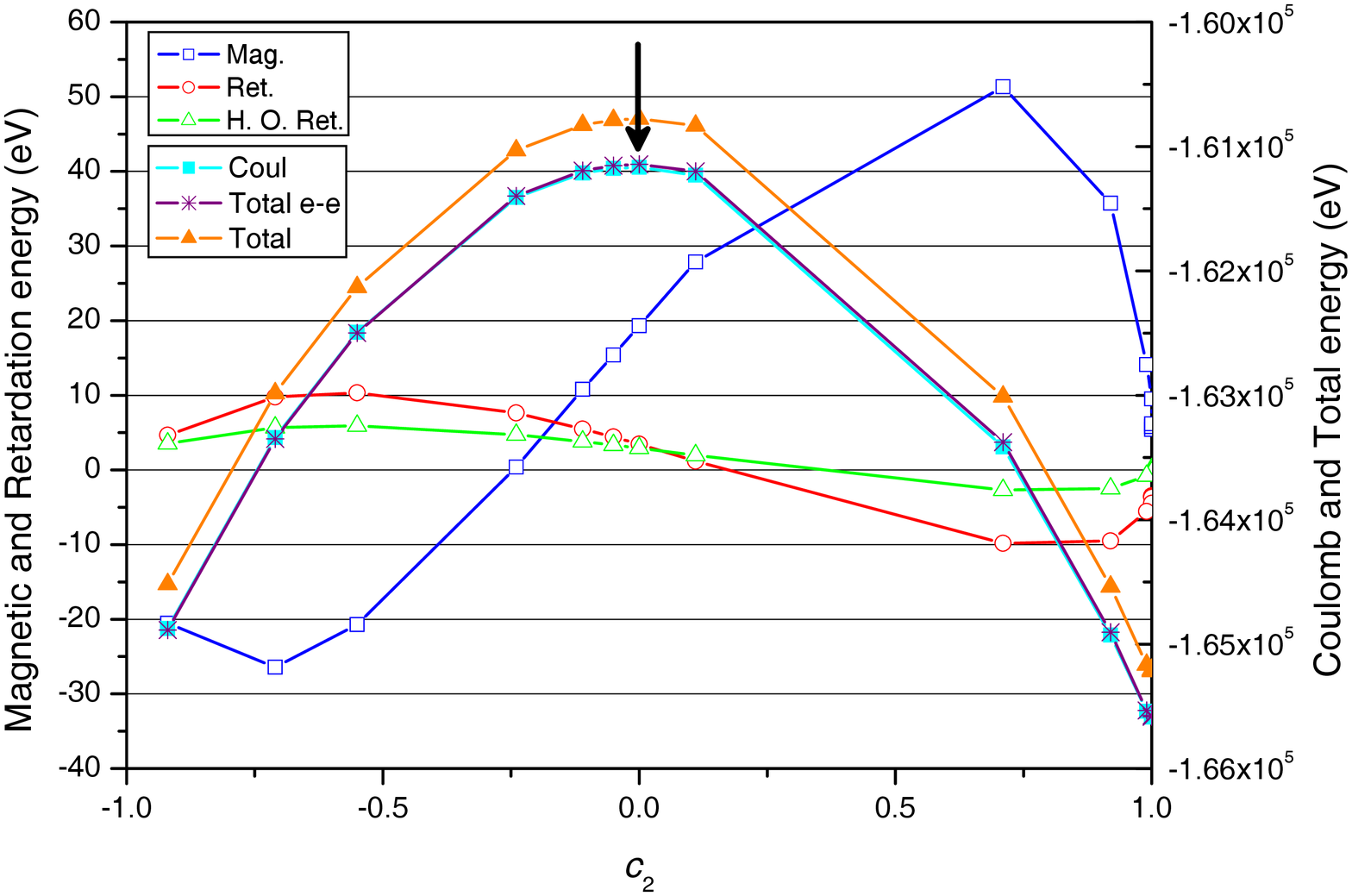}
\caption{Variation of the different contributions to the energy as a function of the mixing coefficient $c_1$, for
the $1s 2p\;^1P_1$ level at $Z=92$. The arrow indicates the position of the $1s 2p \;^1P_1$ energy, at the \emph{maximum}
 around $c_1=0.002$.  \textbf{Left axis:} ``Mag.'': Magnetic energy, Eq. \eqref{eq:magop}. ``Ret.'': Breit retardation,
 Eq. \eqref{eq:breit}. ``H.O. Ret.'': higher-order retardation. \textbf{Right axis:} ``Coul.'': Coulomb
energy Eq. \eqref{eq:coulop}. ``Total e-e'': sum of the 4 preceding contributions.
`` Total'': total level energy including all QED corrections.}
\label{fig:z92-1p1}
\end{figure*}

The use of different form of the electron-electron interaction in the self-consistent field  process has
 a profound qualitative influence on the behavior of variational calculations that  goes beyond changes in energy.
In particular the mixing coefficients between configurations contributing to intermediate coupling are strongly
affected (and thus the values of many operators would be likewise affected).
 For example,  let us examine the very simple case of the $1s2p \;^3P_1$ state in two-electron system.
In a MCDF calculation, intermediate coupling is taken care of by calculating  $\mid 1s2p \;^3P_1\rangle
=c_1\mid 1s2p_{1/2} J=1\rangle+ c_2\mid 1s2p_{3/2} J=1\rangle $. The evolution as a function of the atomic number
$Z$ of the $c_2$ coefficient, is plotted on Fig.\ ~\ref{fig:jj-coupling}, with only the Coulomb interaction, 
or the full Breit interaction made self-consistent. The figure shows clearly that the inclusion of the Breit interaction in the 
SCF process lead to values of $c_2$ that are one order of magnitude lower at high-$Z$ that when only the Coulomb 
interaction is included. It means that the JJ coupling limit is reached much faster. This has some influence
even in the convergence of the calculation: as the exchange potential for the $2p_{3/2}$ orbital is proportional 
to $\left(\frac{c_1}{c_2}\right)^2$, it becomes very large, and the calculation does not converge. This can be traced
back to a negative energy continuum problem. If we use the method described in  Ref.\ ~\cite{ind1995} to solve
for the $2p_{3/2}$, then convergence can be reached, provided the projection operator that suppress 
coupling between positive and negative energy solution of the Dirac equation is used. On Fig. \ref{fig:z40-3p1},
the different contributions to the $1s 2p\;^3P_1$ level energy are plotted at $Z=40$. The minimum, which corresponds to the 
level energy, is obtained for a mixing coefficient (habitually obtained by diagonalization of the Hamiltonian
 matrix), $c_2=0.104$. The shape of the magnetic and retardation energy contribution, as observed on the figure,
shows clearly that the curve used to find the minimum (which represents the sum of the contribution of
the mean values of the operators in Eqs. \eqref{eq:coulop}, Eqs. \eqref{eq:magop}, and \eqref{eq:breit}) is
shifted to the left compared to the pure Coulomb contribution. This explain why the mixing coefficients gets much
smaller when using the Breit interaction in place of the Coulomb interaction in the SCF process.
At high-$Z$ this lead to an extra difficulty to achieve convergence: as can be seen in Figs.  \ref{fig:z92-3p1}
and \ref{fig:z92-1p1}, the minimum corresponding to the $1s 2p\;^3P_1$ state (the one lower in energy) is very
 close to $c_2=0$. It thus sometimes happens during the convergence, that $c_2$ changes sign, leading to very tedious
tuning of the convergence process. This is even worse for the $1s 2p\;^1P_1$ state, because one is trying to reach the
maximum energy. In that case the oscillation of the $c_1$ coefficient around zero are impossible to damp.
Obviously, such problems will slowly disappear when going to neutral systems. For example, in neutral nobelium ($Z=102$)
$c_2$ changes only from 0.47250012 to 0.47316712.


\section{Relativistic and QED effects on transition energies and probabilities}
\label{sec:res}

\subsection{Beryllium isoectronic sequence correlation}
\label{sec:belike}
\begin{figure}[htbp]
\centering
\includegraphics[width=\columnwidth]{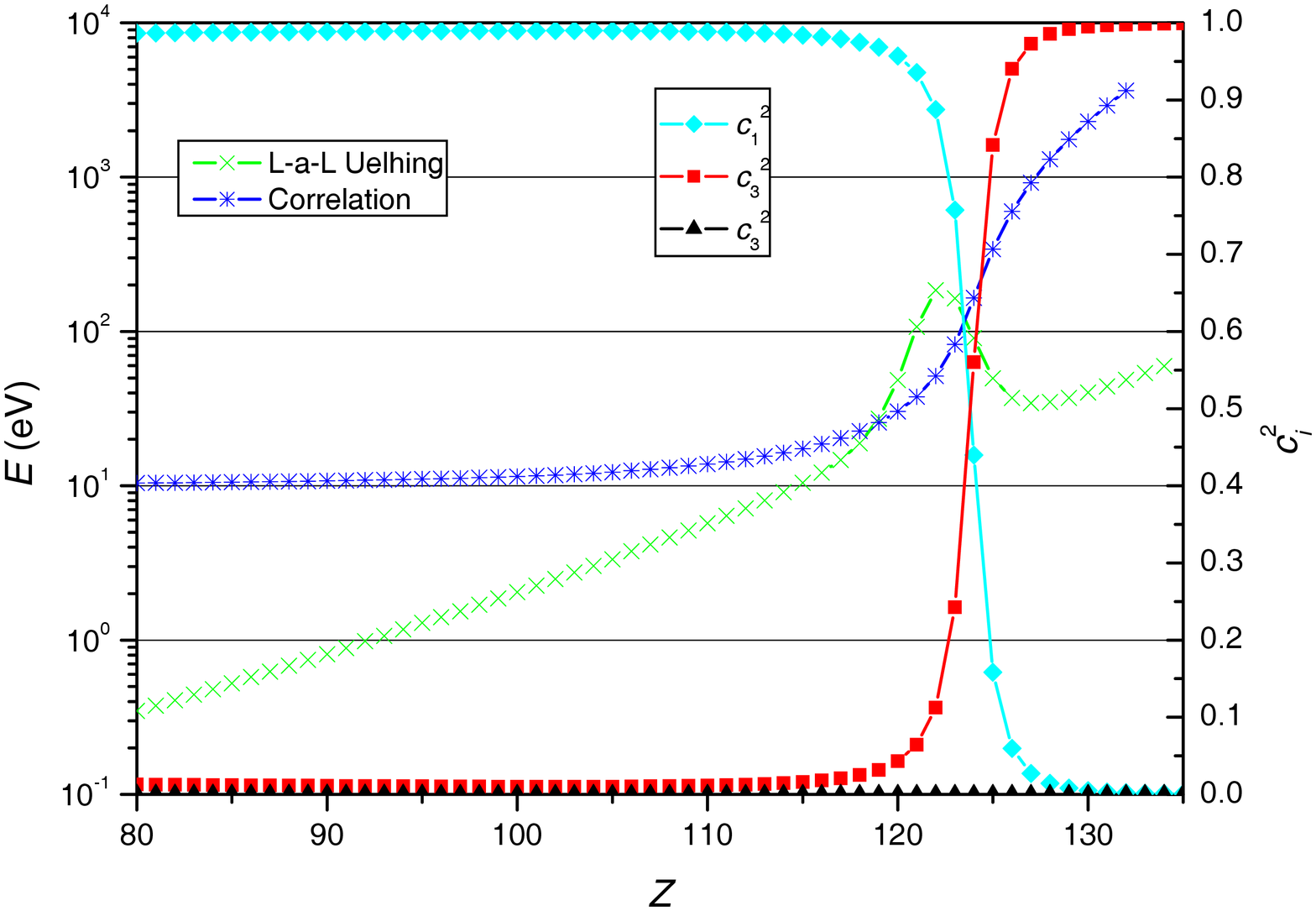}
\caption{Loop-after-loop Uehling contribution to berylliumlike ions ground state energy (changed of sign),
 obtained by including the Uehling potential 
in the SCF. The intrashell correlation energy is also plotted (left axis), as well as the square of
 the mixing coefficients of the $1s² 2s² J=0$, $1s² 2p²_{1/2} J=0$ and  
$1s² 2p²_{3/2} J=0$ configurations (right axis).}
\label{fig:belike}
\end{figure}

It is interesting to investigate simple many-body system, that can be calculated accurately, to see which kind of highly relativistic effects
can be expected in the limit $Z\alpha \to 1$. In that sense the beryllium isoelectronic sequence is an interesting model case, as it exhibit
a very strong intrashell coupling between the $1s^2 2s^2 J=0$ and $1s^2 2p_{1/2}^2 J=0$ configurations, which are almost degenerate
in energy. We thus calculated all contributions to the energy, with and without including the vacuum polarization 
and Breit interaction in the SCF process. From this we could deduce the loop-after-loop Uehling contribution 
to the total energy, and the intrashell correlation. Most quantities contributing to the total energy do not exhibit any specific
behavior when  $Z\alpha \to 1$. However, a major changes in behavior of the system occurs around $Z=125$ as shown
in Fig.\ ~\ref{fig:belike}. One can see that the ground state, which is $1s^2 2s^2 J=0$ at lower $Z$, becomes
 $1s^2 2p_{1/2}^2 J=0$. This can be seen on the mixing coefficients as plotted on Fig.\ ~\ref{fig:belike}.
 This translates into a strong increase in the loop-after-loop vacuum polarization contribution.
Obviously, if we were able to evaluate other second-order QED calculation than loop-after-loop vacuum polarization, including off-diagonal two-electron self-energy 
matrix elements for quasi-degenerate state, following recent work on heliumlike systems \cite{lis2001,lasm2001,asyp2005}, there could be more
unexpected effects to observe.

\begin{figure}[htbp]
\centering
\includegraphics[width=\columnwidth]{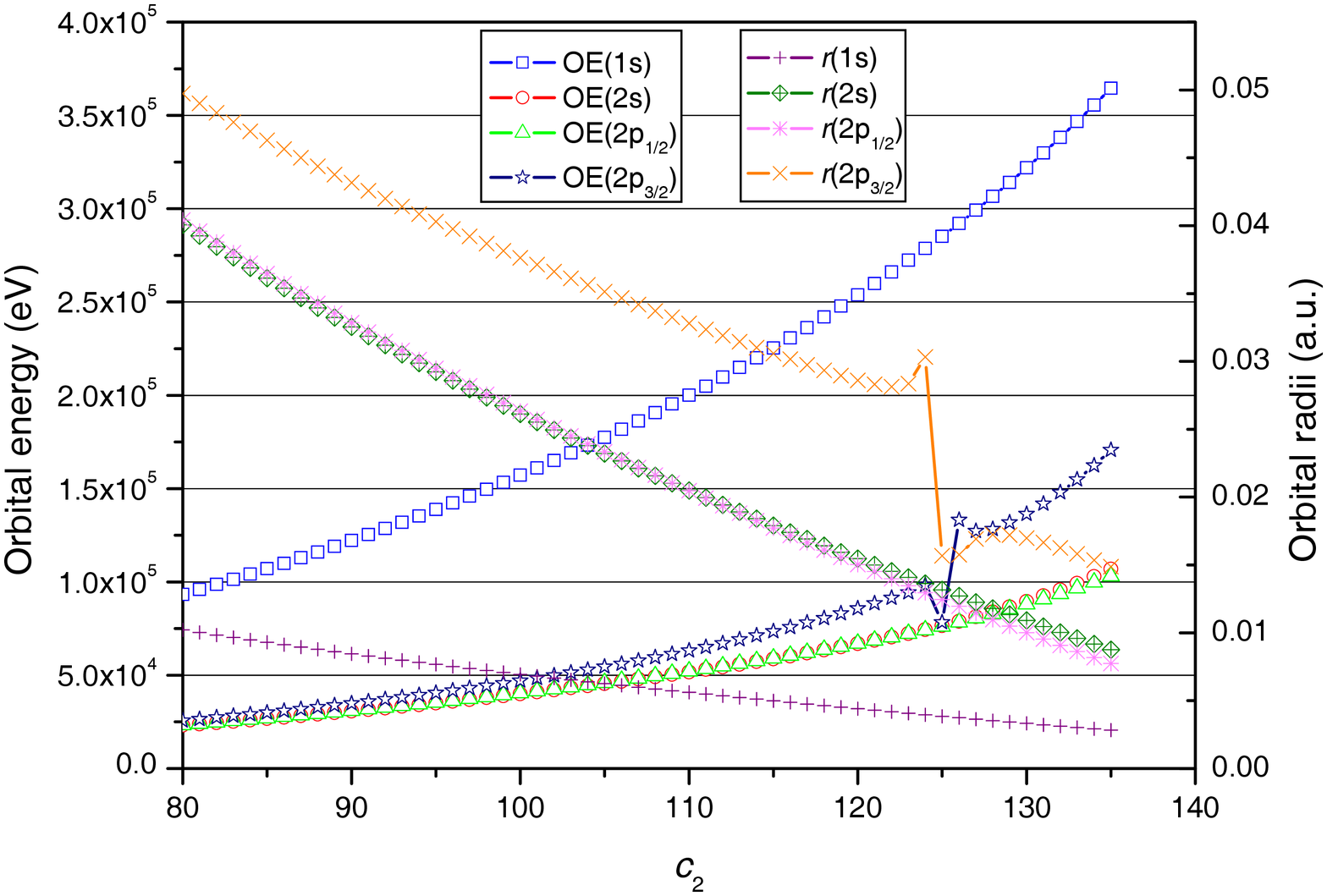}
\caption{Orbital radii (right axis) and one-electron energies (left axis) for Be-like ions}
\label{fig:belike-oe}
\end{figure}

 While not displaying such a feature, the total correlation energy increases strongly, reaching up to 3.6~keV.
One can observe effects on other properties of the atom, like orbital energies and mean orbital radius.
Figure \ref{fig:belike-oe} shows that, in the same atomic number range when the ground state changes of structure,
 the $2p_{3/2}$ orbital radius and energy exhibit  a very strong change. The behavior of the ground state must be connected 
to the fact that the small component of $2p_{1/2}$ orbitals, as can be seen from Eq.\ ~\eqref{eq:dir-wf},
 has a $s$ behavior, and the ratio between small and large component is of order $Z\alpha$. 
It is thus understandable that  such effects could occurs when $Z\alpha \to 1$. It should be noted that this effects happens even with 
a pure Coulomb electron-electron interaction, which is one more proof it is only connected with the behavior of the one-electron wavefunctions.
  We investigated the similar case of 
the magnesiumlike sequence, which exhibit strong intrashell coupling between the $3s$, $3p$ and $3d$ orbitals, but
we could not observe any  effect on energies in this range of $Z$. However the [Ne]$3p_{1/2}^2$ mixing coefficient started
to increase faster around $Z=128$, but convergence problems prohibited us to investigate higher $Z$.

\subsection{Relativistic correlations on the neon isoelectronic sequence}
\label{sec:neonlike}
\begin{figure}[htbp]
\centering
\includegraphics[width=\columnwidth]{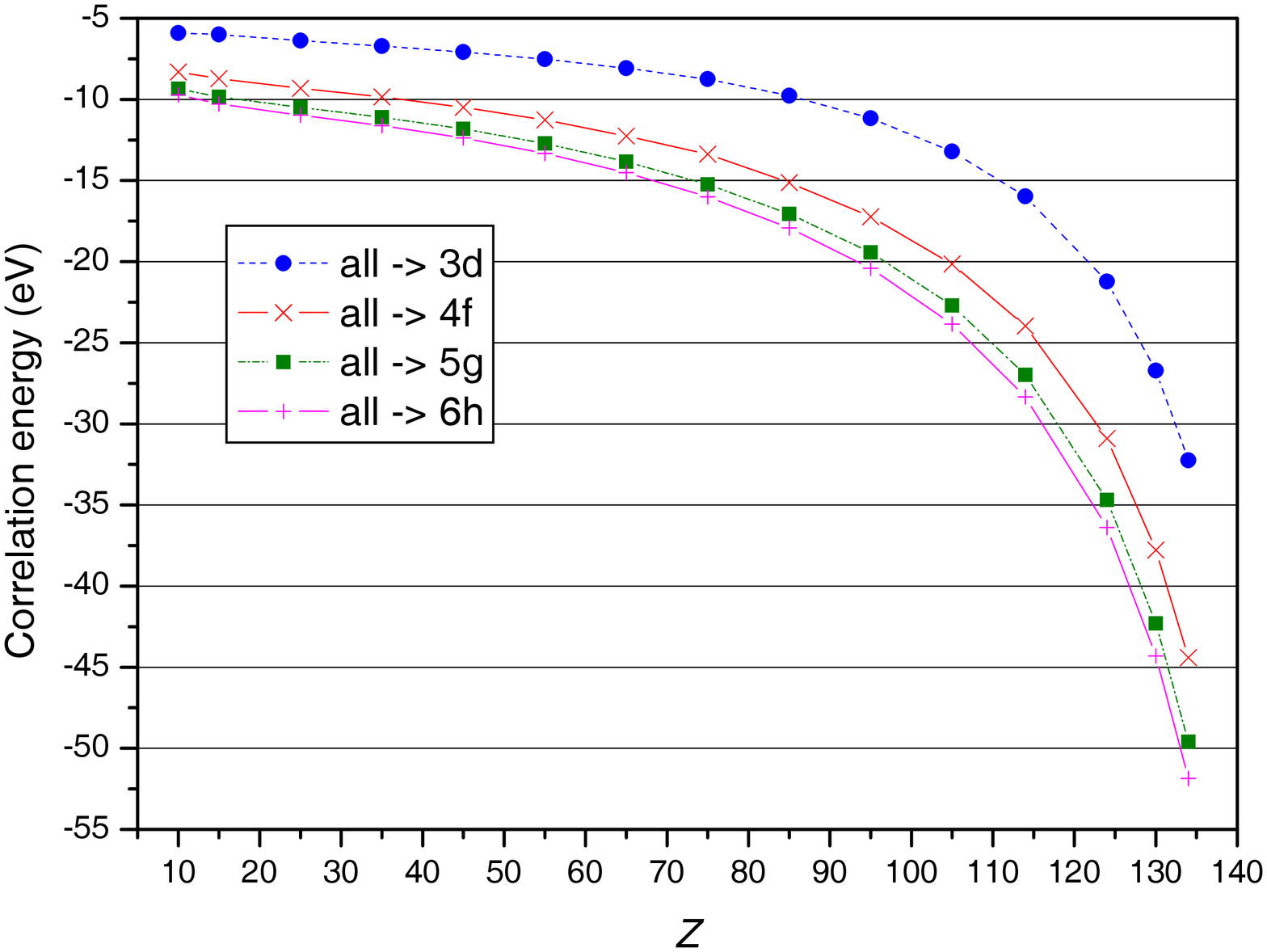}
\caption{Total correlation energy for neonlike ions, with the Breit interaction included in the SCF, as a function
of the most excited orbital included in the basis set.}
\label{fig:neonlike}
\end{figure}

Calculating completely correlated energies can be performed only on relatively small systems. 
Neonlike ions, with 10 electrons are a small enough system that can be calculated with rather large basis sets.
We have extended the calculation performed in Ref.\ ~\cite{srmp2006} to superheavy elements ($Z=134$).
All 10 electrons are excited to all virtual orbitals up to a maximum $n\ell$, that is varied from $3d$ to $6h$.
The results are presented in Table \ref{tab:neonlike} and plotted on Fig.\ ~\ref{fig:neonlike}.
The trend found up to $Z=94$ in  Ref.\ ~\cite{srmp2006} extend smoothly to larger $Z$, but is
enhanced. The total correlation energy becomes very large. It doubles when going from $Z=95$ to $Z=134$, the 
highest $Z$ for which convergence could be reached. The speed of convergence as a function of $n\ell$ does not
change with increasing $Z$.

\begin{table}
\begin{center}
\caption{Total correlation energy for neonlike ions, with the Breit interaction included in the SCF, as a function
of the most excited orbital included in the basis set.
\label{tab:neonlike} }
\begin{tabular}{ldddd}
\hline
\hline
 \multicolumn{1}{c}{$Z$} &\multicolumn{1}{c}{all $\to 3d$} &
\multicolumn{1}{c}{ all $\to 4f$} & \multicolumn{1}{c}{all $\to 5g$} & \multicolumn{1}{c}{all $\to 6h$} \\
\hline
10 & -5.911 & -8.306 & -9.339 & -9.709 \\
15 & -5.989 & -8.712 & -9.838 & -10.280 \\
25 & -6.374 & -9.310 & -10.494 & -10.967 \\
35 & -6.710 & -9.850 & -11.099 & -11.609 \\
45 & -7.074 & -10.482 & -11.816 & -12.372 \\
55 & -7.515 & -11.269 & -12.710 & -13.322 \\
65 & -8.067 & -12.260 & -13.833 & -14.511 \\
75 & -8.752 & -13.375 & -15.247 & -16.007 \\
85 & -9.772 & -15.119 & -17.056 & -17.916 \\
95 & -11.160 & -17.231 & -19.429 & -20.415 \\
105 & -13.205 & -20.146 & -22.700 & -23.853 \\
114 & -15.975 & -23.966 & -26.974 & -28.332 \\
124 & -21.222 & -30.897 & -34.694 & -36.389 \\
130 & -26.714 & -37.777 & -42.304 & -44.300 \\
134 & -32.248 & -44.412 & -49.598 & -51.860 \\
\hline
\hline
\end{tabular}
\end{center}
\end{table}


\subsection{Transition  energies and probabilities in nobelium and element 118 (eka-radon) }
\label{sec:nobelium}
In this section we study the different contributions to the energy and transition probabilities of nobelium ($Z=102$) and eka-radon ($Z=118$). 
Nobelium is the next candidate for measurement of its first excited levels transition energy by laser spectroscopy.
Large scale calculations have been performed recently \cite{fri2005}. Here we examine a number of corrections
not considered in Ref.\ ~\cite{fri2005}. For the $7s^2\;^1S_0$ ground state, we did a MCDF calculation taking into account all single and double
excitations, except the one corresponding to the Brillouin theorem, from the $5f$ and $7s$ shells to the $7p$, and $6d$ shells (48 jj configurations).
For the excited states we included excitations from  $5f$, $7s$ and $7p$ shells to the $7p$ and $6d$ ones. This leads to 671 jj configurations for
the $7s 7p \;^3P_1$ and $7s 7p \;^1P_1$ states and 981 for $7s 7p \;^3P_2$. The results of this calculation are presented on Table ~\ref{tab:nobel-trans-en}
 for the transition energies and  on Table ~\ref{tab:nobel-trans-prob} for transition probabilities.
It is clear from these two tables that many contributions that are important for level energies are completely negligible for transition probabilities in this case.
This is not true for highly charged ions, even at much lower $Z$.
In particular two-loop QED corrections, even though many of them have not been calculated from $n>2$, should remain completely negligible, since they will be
of the same order of magnitude as the loop-after-loop and Källèn and Sabry contributions. The self-energy screening is almost exactly compensating the self-energy, but leaves
a contribution that should be visible if one can calculate correlation well enough. At the present level of accuracy, calculation of the pure Coulomb correlation is the real challenge
as it requires considerable effort on the size of configuration space.

The transition probabilities on Table  \ref{tab:nobel-trans-prob} have been evaluated with different approximation, using theoretical energies. Here the choice
of the wavefunction optimization technique has a sizeable effect. It remains small compared to correlation, but still at the level of 0.2\%.
We also investigated the effect of using fully relaxed orbitals on both initial and final state. In particular we checked at the Dirac-Fock level of approximation,
what is the order of magnitude of taking into account non-orthogonality between initial and final state orbitals. We found 2.2\% for the $^3P_1 \to \;^1S_0$ transition,
0.014\% for the $^1P_1 \to \;^1S_0$ transition and -0.025\% for the $^3P_2 \to \;^1S_0$. For calculations between correlated wavefunctions, it is anyway important to evaluate 
the matrix elements for off-diagonal operators, taking account non-orthogonalities  between initial and final state orbitals, as it can change dramatically the contribution
of a given CSF, since overlaps between correlations orbitals in initial and final states can be very different from either one or zero.

As another example we have also calculated several transition energies and rates for element 118, which are displayed in Table ~\ref{tab:elem118}.
 For the most intense $7p^5 7d \to 7s^2 7p^6$ transitions and for the $7p^5 8s \to 7s^2 7p^6$ transitions we did a MCDF calculation,
 taking into account all single and double excitations,
 (except the ones corresponding to the Brillouin theorem) from the $7s$ and $7p$ shells to the $7d$ and $6f$ shells for the $7p^6\;^1S_0$ ground state (38 jj configurations)
 and for the $7p^5 7d$ excited states (657 jj configurations), and from the $7p$ and $8s$ shells to the $7d$ and $6f$ shells for the $7p^5 8s$
 excited states (151 jj configurations).

\begin{figure}[htbp]
\centering
\includegraphics[width=\columnwidth]{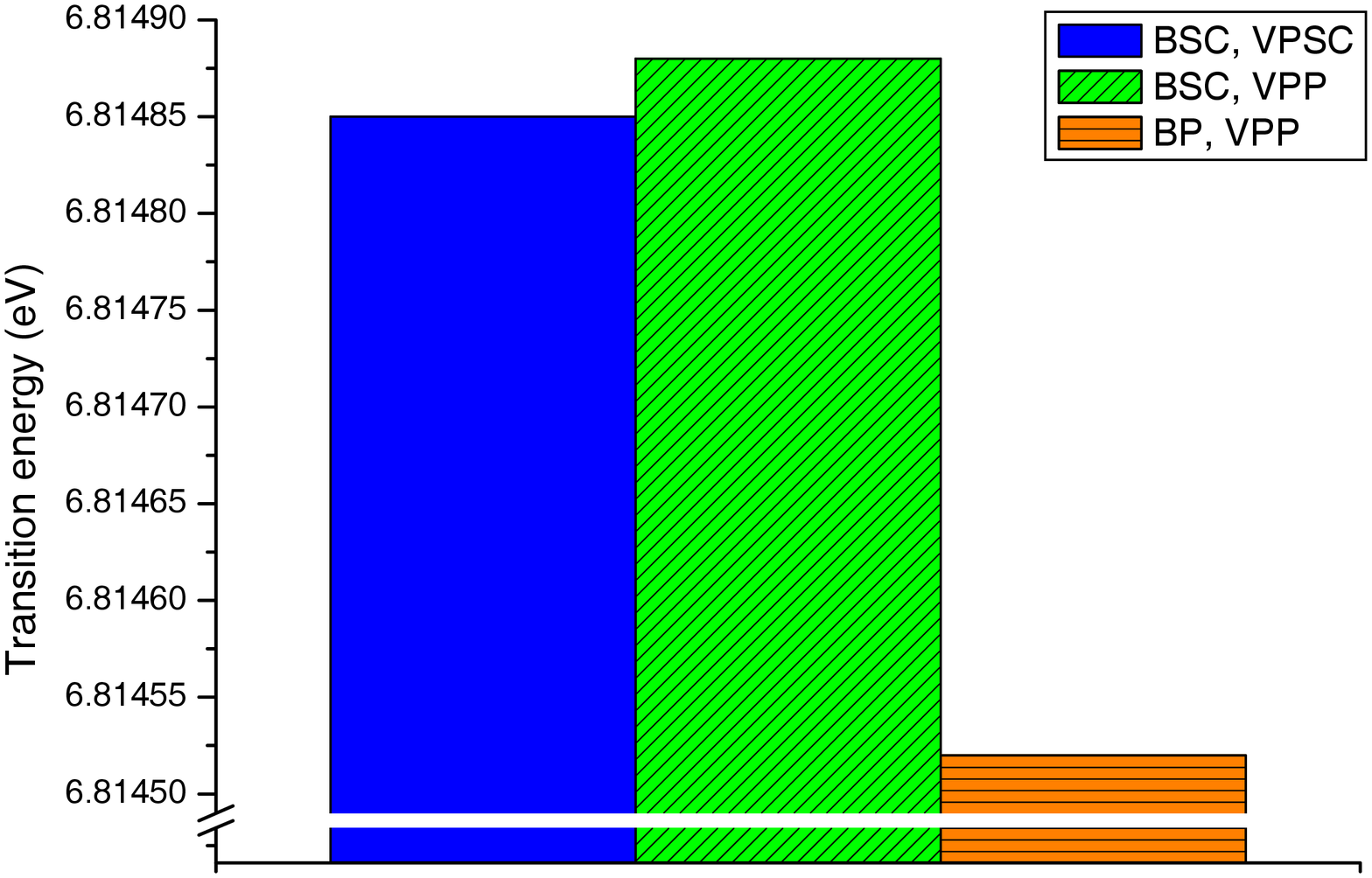}
\caption{Changes in the $7p^5 8s\;^3P_1\to 7s^2 7p^6\;^1S_0$  transition energy with respect to the approximation made, for element 118. BSC: Breit self-consistent. VPSC: Uelhing potential
self consistent. BP: Breit in perturbation. VPP: Uelhing vacuum polarization in perturbation.}
\label{fig:eka-radon}
\end{figure}

The transition energy was calculated for some of these transitions, with and without including the vacuum polarization and Breit interaction in the SCF process, 
and we concluded that the transition energy is not significantly affected by the inclusion of these interactions in the SCF. This is illustrated 
on Fig.\ ~\ref{fig:eka-radon} that shows the transition energy values for $7p^5 7d\;^3P_1$ transition.
%
%
\begin{table*}
\begin{center}
\caption{Transition energies for the lower energy levels of nobelium (eV). Coul.: DF Coulomb energy. Mag. (pert), Ret. (pert.), Higher order ret. (pert.):
 Contribution of the Magnetic, Breit retardation and higher-order retardation  in first order of perturbation.
All order Breit: effect of including the full Breit interaction in the SCF. Coul. Corr.: Coulomb correlation. Breit Corr.:Contribution
of all Breit terms to the correlation energy. Self-energy (FS): self-energy and finite size correction. Self-energy screening: Welton approximation to
self-energy screening. Vac. Pol. (Uehling): mean value of the Uehling potential (order $\alpha (Z\alpha)$). VP (muons, Uehling): vacuum polarization due to muon loops. 
VP Wichman and Kroll: correction to the Uehling potential (order $\alpha (Z\alpha)^3$). Loop-after-loop V11: iterated Uehling contribution to all orders. 
VP (Källèn et Sabry): two-loop contributions to vacuum polarization. Other 2nd order QED: sum of two-loop QED corrections not accounted for in the two previous one. Recoil:
sum of lowest order recoil corrections (see, e.g., \cite{mat2000}). The number of digits presented in the table  is not physically significant, but is necessary to show the size of some
contributions. The physical accuracy is not better than 1  digits.
\protect\label{tab:nobel-trans-en} }
\begin{tabular}{lf{5}f{5}f{5}}
\hline
\hline
Contribution &   \multicolumn{1}{c}{$^3P_1 \to \;^1S_0$} & \multicolumn{1}{c}{$^1P_1 \to \;^1S_0$} & \multicolumn{1}{c}{$^3P_2 \to \;^1S_0$} \\
\hline                              
Coul. & 1.69909 & 3.20051 & 2.19931 \\
Mag. (pert) & 0.00070 & -0.00420 & -0.00347 \\
Ret. (pert.) & -0.00048 & 0.00005 & -0.00058 \\
Higher order ret. (pert.) & -0.00119 & 0.05048 & -0.00231 \\
All order Breit (pert) & 0.00002 & -0.05877 & 0.00003 \\
Coul. Corr. & 0.59560 & 0.60414 & 0.55545 \\
Breit Corr. & -0.00208 & -0.00229 & -0.00194 \\
Self-energy (FS) & -1.60884 & -1.72025 & -1.76085 \\
Self-energy screening & 1.59030 & 1.72940 & 1.74228 \\
Vac. Pol. (Uehling) & 0.00610 & 0.00435 & 0.00648 \\
VP (muons. Uehling) & 0.00000 & 0.00000 & 0.00000 \\
VP Wichman and Kroll & -0.00030 & -0.00019 & -0.00033 \\
Loop-after-loop Uehl. & 0.00002 & 0.00012 & 0.00003 \\
VP (Källèn and Sabry)&  0.00004 & 0.00002 & 0.00005 \\
Other 2nd order QED & 0.00000 & 0.00000 & 0.00000 \\
Recoil & -0.00082 & -0.00081 & -0.00082 \\
\hline      
Total	&	2.28	&	3.80	&	2.73	\\
Ref. \cite{fri2005} II	&	2.34	&	3.49	&		\\
Ref. \cite{fri2005} I	&	2.60	&	3.36	&		\\
\hline                              
\hline                              
\end{tabular}              
\end{center}              
\end{table*}              
\begin{table*}
\begin{center}
\caption{Effect of correlation, Breit interaction and all order vacuum polarization on transition probabilities of nobelium. DF: Dirac-Fock. VP: vacuum polarization.
ex.: excitation
\protect\label{tab:nobel-trans-prob} }
\begin{tabular}{ld|d|d}
\hline
\hline
initial & \multicolumn{1}{c|}{$7s 7p \;^{2S+1}P_J$ DF}   & \multicolumn{1}{c|}{$7s 7p \;^{2S+1}P_J$ ex. $7s$, $7p \to 7p$, $6d$ }   & \multicolumn{1}{|c}{$7s 7p \;^{2S+1}P_J$ ex. $5f$, $7s$, $7p \to 7s$, $7p$, $6d$ }   \\
final &   \multicolumn{1}{c|}{$\to 7s^2 \;^1S_0$ DF} & \multicolumn{1}{c|}{$\to 7s^2 \;^1S_0$ ex. $7s \to 7p$, $6d$ }   & \multicolumn{1}{|c}{$\to 7s^2 \;^1S_0$ ex. $7s \to 7p$, $6d$ }   \\
\hline
 $7s 7p\;^3P_1 \to 7s^2\;^1S_0$ (E1) & & &           \\
Breit Pert.; VP. Pert. & 1.071 × 10^{6} & 3.520 × 10^{6} & 1.947 × 10^{6} \\
Breit Pert.; VP SC & 1.073 × 10^{6} & 3.526 × 10^{6} & 1.954 × 10^{6} \\
Breit SC; VP Pert. & 1.062 × 10^{6} & 3.501 × 10^{6} & 1.920 × 10^{6} \\
Breit SC; VP SC & 1.064 × 10^{6} & 3.507 × 10^{6} & 1.927 × 10^{6} \\
\hline          
 $7s7p\;^1P_1 \to 7s^2\;^1S_0$ (E1) & & &       \\
Breit Pert.; VP. Pert. & 2.705 × 10^{8} & 3.500 × 10^{8} & 3.680 × 10^{8} \\
Breit Pert.; VP SC & 2.697 × 10^{8} & 3.559 × 10^{8} & 3.738 × 10^{8} \\
Breit SC; VP Pert. & 2.711 × 10^{8} & 3.509 × 10^{8} & 3.686 × 10^{8} \\
Breit SC; VP SC & 2.704 × 10^{8} & 3.568 × 10^{8} & 3.743 × 10^{8} \\
\hline          
$7s7p\;^3P_2 \to 7s^2\;^1S_0$ (M2) & & &         \\
Breit Pert.; VP. Pert. & 1.717 × 10^{-4} & 5.187 × 10^{-4} & 5.492 × 10^{-4} \\
Breit Pert.; VP SC & 1.713 × 10^{-4} & 5.173 × 10^{-4} & 5.477 × 10^{-4} \\
Breit SC; VP Pert. & 1.721 × 10^{-4} & 5.200 × 10^{-4} & 5.513 × 10^{-4} \\
Breit SC; VP SC & 1.717 × 10^{-4} & 5.187 × 10^{-4} & 5.499 × 10^{-4} \\

\hline                              
\hline                              
\end{tabular}              
\end{center}              
\end{table*}              

%
%
\begin{table*}
\begin{center}
\caption{Transition energies and transition rates to the ground state $5f^{15} 6d^{10} 7s^2 7p^6\;^1S_0$ of eka-radon (element 118).
The transition values from states labeled with $^c$ were calculated with correlation up to $6f$, as well as the ground state.
\protect\label{tab:elem118} }
\begin{tabular}{lccdd}
\hline
\hline
\multicolumn{3}{c}{Initial State}     &\multicolumn{1}{c}{ Transition Energy} &
 \multicolumn{1}{c}{Transition Rate}    \\
 &  &  & \multicolumn{1}{c}{(eV)} & \multicolumn{1}{c}{(s$^{-1}$)}    \\
\hline            
\multicolumn{1}{c}{$5f^{15}$ $6d^{10}$ $7s^2$ $7p^5$ $7d$} & \multicolumn{1}{c}{$^3D_1 + ^3P_1$ }& \multicolumn{1}{c}{$^c$}
 & 10.43 & 9.86  × 10^{ 8 } \\
 & $^3P_1$ & $^c$ & 6.81 & 5.53 × 10^{ 7 } \\
 & $^3D_1 + ^1P_1$ & $^c$ & 7.2 & 9.98 × 10^{ 6 } \\
 & $^3P_2$ &  & 6.08 & 1.41 × 10^{ -2 } \\
 & $^3F_2$ &  & 16.95 & 1.11 × 10^{ -2 } \\
 & $^3F_3$ &  & 6.15 & 1.30 × 10^{ -3 } \\
 & $^1D_2$ &  & 6.25 & 1.10 × 10^{ -3 } \\
 & $^3D_3$ &  & 6.22 & 7.52 × 10^{ -4 } \\
 & $^3F_4$ &  & 6.01 & 3.91 × 10^{ -14 } \\
 & $^3P_0$ & $^c$ & 6.64 &     \\
\multicolumn{1}{c}{$5f^{15}$ $6d^{10}$ $7s^2$ $7p^5$ $8s$ }& \multicolumn{1}{c}{$^1P_1$} & \multicolumn{1}{c}{$^c$} & 4.73 & 2.04 × 10^{ 8 } \\
 & $^3P_2$ & $^c$ & 4.3 & 2.04  × 10^{ -3 } \\
\hline       
\hline       
\end{tabular}       
\end{center}       
\end{table*}       
%
%

\section{Relativistic and QED effects on Landé factors, atomic radii  and electronic densities}
\label{sec:land-rad}
Although transition energies and probabilities, as well as ionization potential, are important quantities,
it is interesting to study relativistic and QED effects on other atomic parameters, like Landé 
$g$-factors, atomic radii  and electronic densities. Landé factors, which define the strength of the coupling
of an atom to a magnetic field, can help characterize a level. 

In some experiments concerning superheavy elements, singly ionized atoms are drifted in a gas cell, under 
the influence of an electric field. The drift speed can be related, in first approximation, to the 
charge distribution radius of the ion \cite{bdhk2005,slh2007}. In this context, it can also be interesting to look at the atomic density, and
see how it is affected by relativistic and QED effects. In that case, though, we can only get a feeling
of this effect by comparing densities calculated with and without the Breit interaction self-consistent, or
with and without the Uehling potential included in the differential equation. There is currently no formalism
that would enable to account for changes in the wavefunction related to the self-energy.

We define the radial electronic density as
\begin{eqnarray}
r^2\rho \left(r\right) &=&\int d\Omega r^2 \rho \left(r,\Omega \right) \nonumber  \\
&=&  \sum_{i \in \mathrm{occ. orb.}} \varpi_i \left(P_i\left(r\right)^2+Q_i\left(r\right)^2\right) ,
\label{eq:el-dens}
\end{eqnarray}
where $P$ and $Q$ are defined in Eq.\ ~\eqref{eq:dir-wf}, and  the $\varpi_i$ are the orbital effective occupation numbers,
\begin{equation}
 \varpi_i = \sum_{j} c_{j}^2 n_j^i ,
\end{equation}
where the sum extend over all configuration containing orbital $i$, and $n_j^i$ is the occupation number of this orbital 
in the configuration $j$. The density is normalized to $\int_{0}^{\infty} dr r^2\rho \left(r\right) =N_{e}$, where $N_{e}$ is the number
of electrons in the atom or ion.

\begin{figure}[htbp]
\centering
\includegraphics[width=\columnwidth]{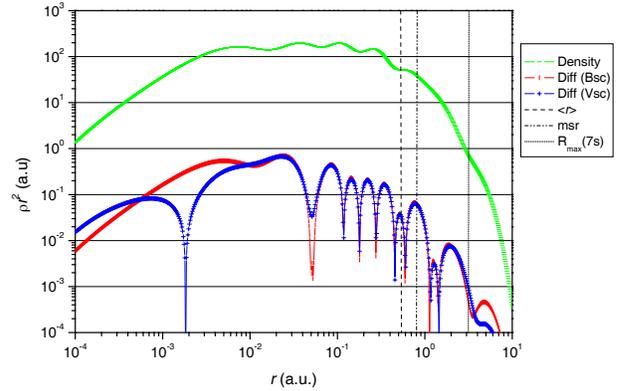}
\caption{Total electronic density of Fm$^{+}$. The vertical lines represents, from left to right the
mean radius, the mean spherical radius (``rms'') and the mean radius of the outermost orbital ($7s$).
``Density'': total electronic density Eq. \eqref{eq:el-dens}. ``Diff (Bsc)'': absolute value of the variation in the density due to the inclusion
of the Breit interaction in the SCF. ``Diff (Vsc)'': absolute value of the variation  in the density due to the inclusion of the Uehling potential
in the differential equations (NB: the dip in the curve correspond to sign changes of the correction). }
\label{fig:fermium-pdens}
\end{figure}

The effect of the Breit interaction and vacuum polarization on the charge density of the ground state of Fm$^+$ ([Rn]­$5f^{12} 6s$) is shown 
on Fig.\ ~\ref{fig:fermium-pdens}. The inclusion of both contributions leads to local changes of around 1\% in 
the charge density. It is rather unexpected that both contribution extend their effects way pass their range,
in particular for the vacuum polarization potential, which has a very small contribution for $r>1/\alpha = 0.0073$~a.u.


\begin{figure}[htbp]
\centering
\includegraphics[width=\columnwidth]{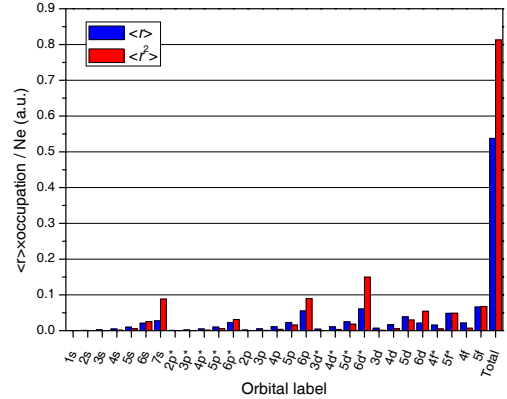}
\caption{Contribution to the mean radius $\langle r_{\mathrm{at.}}^{(p)}\rangle$ to Sg$+$  of individual
values of $\varpi_i  \langle r_{n_i,l_i,j_i}^{(p)}\rangle$ for the different orbitals and $p=1$ and 2, together with the
total values on the right. Stared orbital labels correspond to the orbital with $j=l-\frac{1}{2}$.
\label{fig:sgrad}
}
\end{figure}

\begin{figure}[htbp]
\centering
\includegraphics[width=\columnwidth]{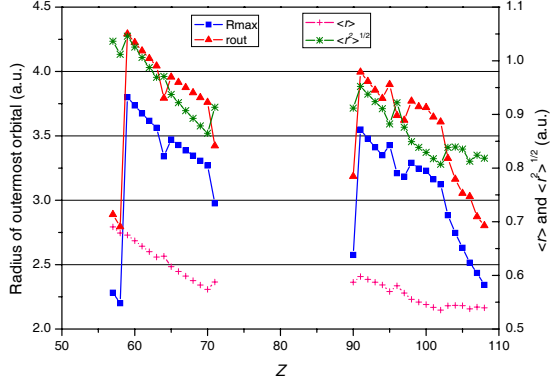}
\caption{Variation of $<r>$, $\sqrt{<r^2>}$ (\textbf{right scale}),  of the radius of the outermost orbital  $<r_{\mathrm{out}}>$ 
 and of the maximum of the charge density of the outer orbital (Rmax. dens.)  (\textbf{left scale} as a function of
the atomic number $Z$.
\label{fig:var-rad}
}
\end{figure}

Charge distribution can be described in a number of ways, and the definition of atomic or ionic radius has
evolved over the years. Slater noticed a long time ago the correlation between the value of the maximum
charge density of the outermost core electron shell and the ionic radius of an atom \cite{sla1964}. Such a definition, however,
does not provide a way to take into account mixing of outer shell in MCDF calculations. Other authors
have chosen either the mean radius of a specific orbital $\langle r_{n,l,j}\rangle $ \cite{sie1997} or weighted mean radius \cite{bil2000}.
Results have been obtained for bohrium and hassium.
It is not at all obvious
either than what is true in crystals must be valid for singly charged ions drifting in a gas.
Here we have evaluated four different quantities and their dependence in the Breit interaction and
vacuum polarization. We have evaluated the position of the maximum density of the outermost orbital, the mean
radius of the outermost orbital $\langle r_{n,l,j}\rangle $, the atom mean radius, and the atom mean spherical radius.
The two first quantities have been tested in detail, but they cannot represent the ionic radius when one has 
a complex outer shell structure or when one calculate correlation. 
The mean radius of the atom is represented as
\begin{eqnarray}
\langle r_{\mathrm{at.}}^{(p)}\rangle  & = &\frac{1}{N_e}\int_{0}^{\infty} dr r^p r^2 \rho \left(r\right) \nonumber \\
                  & = &\frac{1}{N_e} \sum_{i \in \mathrm{occ. orb.}} \varpi_i  \langle r_{n_i,l_i,j_i}^{(p)}\rangle
\end{eqnarray}
for $p=1$, and the mean spherical radius is obtained as $\sqrt{\langle r_{\mathrm{at.}}^{(2)}\rangle}$. Both can be calculated
in a MCDF model, as the $ \varpi_i$ contain both the occupation numbers and the mixing coefficients.
As an example, we have plotted the contribution to $\langle r_{\mathrm{at.}}^{(p)}\rangle$ to Sg$^+$ on Fig.\ ~\ref{fig:sgrad} of individual
values of $\varpi_i  \langle r_{n_i,l_i,j_i}^{(p)}\rangle$ for the different orbitals. Contrary to the individual orbitals quantities
or to the average performed only on the outer shell, like in Ref. ~\cite{jfjd2002}, both quantities have sizeable
contributions from several outer shells. The $7s$, $6p$, $5f$ and $6d$ contribute
significantly to both the mean and the mean spherical atomic radius, but the contribution of the $6d$ orbitals is more dominant
for the mean spherical radius. The results for singly charged ions with
$57 \leq Z \leq 71$ and  $90\leq Z \leq 108$ are presented on Table \ref{table:rad-sing-ion}  and plotted on Fig.\ ~\ref{fig:var-rad}. The comparison between mean radius and mean spherical radius
 with the Breit interaction in the SCF process  with Coulomb values,
shows changes  around 0.04\% to 0.08\%  depending on the atomic number. Self-consistent vacuum polarization has an effect 2 to 4 times smaller, depending on the
element.  We emphasize the fact that we used for the ion the configuration corresponding to the ground state of a Dirac-Fock calculation (i.e., without correlation),
 as given in \cite{risp2004}. For a few elements, the physical ground state configuration, as given for example on the NIST database, is different.

We have used the definition above to evaluate the effect of correlation on the ion radii. The calculation has been 
performed on neutral nobelium ($Z=102$). The results are presented on Table \ref{tab:rad-corr}.
One can see that the radius of maximum density follows exactly the trend of the mean radius of the outer orbital. However the atomic mean radius and mean spherical radius follow
a different trend.

There has been two experiments measuring drift time of singly ionized transuranic elements in gases. One measured the relative velocity of Am$^+$ with respect to
Pu$^+$ \cite{bdhk2005} and an other one  measured the same quantity for Cf$^+$ and Fm$^+$ \cite{sbdk2003}. The results of these experiments are expressed as
\begin{equation}
\Delta r_{A-B}=\frac{r_A-r_B}{r_B}.
\end{equation}
The experiments above provide  $\Delta r_{\mathrm{Am}^+-\mathrm{Pu}^+}$=-$3.1± 1.3$\% and  $\Delta r_{\mathrm{Fm}^+-\mathrm{Cf}^+}$=-2\% respectively, which shows a shrinkage of
Am as compares to Pu and of Fm as compared to Cf.
The comparison between these experiments and the calculation of singly charged ion radii from Table \ref{table:rad-sing-ion} is shown in 
Table \ref{tab:rad-exp}. Although it is difficult to draw any firm conclusion from so few data, it seems that $\sqrt{\langle r_{\mathrm{at.}}^{(2)}\rangle}$
reproduces best the experiment, followed by $\langle r_{\mathrm{at.}}^{(1)}\rangle$. We want to point out that the $<r_{7s}>$ values for neutral atom, from Ref.\ ~\cite{des1973}, or ours, which are
in good agreement gives a -1.5\% change for Am/Pu and -3.0\% for Fm/Cf, compare well with experiment. The $7s$ radius of the singly charged ions, as shown
 in Table  \ref{table:rad-sing-ion}, increases when going from Am to Pu, thus leading to a ratio of the wrong sign, while the $5f$ average radius, or the global atomic radii as
defined here all show the right trend.
The change of behavior for Am, is due to the relative diminution of the $l (l+1)/r^2$ barrier compared to the Coulomb potential as a function of $Z$.
The $5f$ radius reduces then strongly, leading to the change of the structure of the ground configuration, and to the change of the radius of the  $7s$ orbital.


 \begin{table*}
 \begin{center}
 \caption{Singly ionized atom radii (a.u.) for lanthanide $57\leq Z \leq 71$ and actinide and transactinide 
  $ 90\leq Z\leq 108$ \label{table:rad-sing-ion} }
 %
 \scriptsize{
 \begin{tabular}{rp{10mm}lcccccccccc}
 \hline
 \hline
 Z &\multicolumn{1}{c}{ Config.} & \multicolumn{1}{c}{Label} &\multicolumn{1}{c}{ $<r>$}
  &\multicolumn{1}{c}{ $\sqrt{<r^2>}$} 
    & \multicolumn{2}{c}{outer orb.}\\
   &         &       &                           &                                &
  \multicolumn{1}{c}{lab.}    &\multicolumn{1}{c}{$R_{\mbox{max} }$}& $<4f_{5/2}>$ & $<4f_{7/2}>$ & $<5d{3/2}>$ & $<5d_{7/2}>$ & $<6s_{1/2}>$ \\
 \hline
57 &$ 5d^{2}        $&$ \;^{3}F_{2} $& 0,6903 & 1,0365 &$ 5d $& 2,2820 &  &  & 2,8629 & 2,8900 &  \\
58 &$ 4f 5d^{2}     $&$ \;^{4}H_{7/2} $& 0,6789 & 1,0121 &$ 5d $& 2,2005 & 1,0862 & 1,1033 & 2,7568 & 2,7949 &  \\
59 &$ 4f^{3} 6s          $&$ \;^{5}I_{4} $& 0,6750 & 1,0458 &$ 6s $& 3,8004 & 1,0589 & 1,0667 &  &  & 4,2924 \\
60 &$ 4f^{4} 6s          $&$ \;^{6}I_{7/2} $& 0,6646 & 1,0256 &$ 6s $& 3,7359 & 1,0054 & 1,0190 &  &  & 4,2252 \\
61 &$ 4f^{5} 6s          $&$ \;^{7}H_{2} $& 0,6544 & 1,0062 &$ 6s $& 3,6741 & 0,9624 & 0,9796 &  &  & 4,1608 \\
62 &$ 4f^{6} 6s          $&$ \;^{8}F_{1/2} $& 0,6442 & 0,9874 &$ 6s $& 3,6171 & 0,9249 & 0,9392 &  &  & 4,1012 \\
63 &$ 4f^{7} 6s          $&$ \;^{9}S_{4} $& 0,6342 & 0,9692 &$ 6s $& 3,5627 & 0,8920 & 0,8999 &  &  & 4,0438 \\
64 &$ 4f^{7} 5d 6s          $&$ \;^{10}D_{5/2} $& 0,6357 & 0,9708 &$ 6s $& 3,3408 & 0,8218 & 0,8221 & 2,4547 & 2,4846 & 3,7930 \\
65 &$ 4f^{9} 6s          $&$ \;^{7}H_{8} $& 0,6162 & 0,9373 &$ 6s $& 3,4711 & 0,8502 & 0,8541 &  &  & 3,9568 \\
66 &$ 4f^{10} 6s          $&$ \;^{6}I_{17/2} $& 0,6074 & 0,9219 &$ 6s $& 3,4289 & 0,8254 & 0,8345 &  &  & 3,9163 \\
67 &$ 4f^{11} 6s          $&$ \;^{5}I_{8} $& 0,5987 & 0,9070 &$ 6s $& 3,3871 & 0,8022 & 0,8165 &  &  & 3,8758 \\
68 &$ 4f^{12} 6s          $&$ \;^{4}H_{13/2} $& 0,5903 & 0,8926 &$ 6s $& 3,3464 & 0,7804 & 0,7995 &  &  & 3,8359 \\
69 &$ 4f^{13} 6s          $&$ \;^{3}F_{4} $& 0,5820 & 0,8786 &$ 6s $& 3,3070 & 0,7605 & 0,7811 &  &  & 3,7976 \\
70 &$ 4f^{14} 6s          $&$ \;^{2}S_{1/2} $& 0,5739 & 0,8649 &$ 6s $& 3,2699 & 0,7430 & 0,7613 &  &  & 3,7611 \\
71 &$ 4f^{14} 6s^{2}     $&$ \;^{1}S_{0} $& 0,5876 & 0,9133 &$ 6s $& 2,9761 & 0,6922 & 0,7057 &  &  & 3,4229 \\
\hline                         
   &         &       &                           &                                &
   &                 & $<5f_{5/2}>$ & $<5f_{7/2}>$ & $<6d_{3/2}>$ & $<6d_{7/2}>$ & $<7s_{1/2}>$ \\
90 &$ 5f^{2} 6d          $&$ \;^{4}K_{11/2} $& 0.5873 & 0.9117 &$ 6d $& 2.5751 & 1.96004 & 1.95812 & 3.16141 & 3.18613 &  \\
91 &$ 5f^{2} 6d 7s          $&$ \;^{5}K_{5} $& 0.5977 & 0.9521 &$ 7s $& 3.5491 & 1.54181 & 1.57544 & 2.85704 & 2.9741 & 3.99617 \\
92 &$ 5f^{3} 6d 7s          $&$ \;^{6}L_{11/2} $& 0.5926 & 0.9377 &$ 7s $& 3.4779 & 1.42706 & 1.44552 & 2.78857 & 2.86351 & 3.92351 \\
93 &$ 5f^{4} 6d 7s          $&$ \;^{7}L_{5} $& 0.5873 & 0.9239 &$ 7s $& 3.4116 & 1.3478 & 1.36699 & 2.72666 & 2.79424 & 3.85537 \\
94 &$ 5f^{5} 6d 7s         $&$ \;^{8}K_{7/2} $& 0.5821 & 0.9107 &$ 7s $& 3.3498 & 1.28511 & 1.30541 & 2.67874 & 2.73928 & 3.7917 \\
95 &$ 5f^{7} 7s       $&$ \;^{9}S_{4} $& 0.5699 & 0.8822 &$ 7s $& 3.4279 & 1.27013 & 1.29869 &  &  & 3.8992 \\
96 &$ 5f^{7} 7s^{2}         $&$ \;^{8}S_{7/2} $& 0.5805 & 0.9217 &$ 7s $& 3.2104 & 1.16578 & 1.18716 &  &  & 3.65864 \\
97 &$ 5f^{8} 6d 7s         $&$ \;^{9}G_{7} $& 0.5670 & 0.8761 &$ 7s $& 3.1840 & 1.14449 & 1.16614 & 2.62581 & 2.688 & 3.62195 \\
98 &$ 5f^{10} 7s          $&$ \;^{6}I_{17/2} $& 0.5553 & 0.8493 &$ 7s $& 3.2905 & 1.14462 & 1.18171 &  &  & 3.77066 \\
99 &$ 5f^{11} 7s          $&$ \;^{5}I_{8} $& 0.5503 & 0.8385 &$ 7s $& 3.2461 & 1.10705 & 1.14777 &  &  & 3.72772 \\
100 &$ 5f^{12} 7s          $&$ \;^{2}H_{11/2} $& 0.5456 & 0.8293 &$ 7s $& 3.2294 & 1.07281 & 1.11537 &  &  & 3.72114 \\
101 &$ 5f^{13} 7s          $&$ \;^{3}F_{4} $& 0.5403 & 0.8174 &$ 7s $& 3.1625 & 1.0417 & 1.08491 &  &  & 3.64662 \\
102 &$ 5f^{14} 7s          $&$ \;^{2}S_{1/2} $& 0.5353 & 0.8071 &$ 7s $& 3.1235 & 1.01352 & 1.05406 &  &  & 3.6087 \\
103 &$ 5f^{14} 7s^2          $&$ \;^{1}S_{0} $& 0.5435 & 0.8385 &$ 7s $& 2.8837 & 0.95846 & 0.990419 &  &  & 3.32578 \\
104 &$ 5f^{14} 6d 7s^2          $&$ \;^{2}D_{3/2} $& 0.5441 & 0.8392 &$ 7s $& 2.7464 & 0.917421 & 0.941578 & 2.33571 &  & 3.16377 \\
105 &$ 5f^{14}  6d^2   7s^2      $&$ \;^{3}F_{2} $& 0.5436 & 0.8355 &$ 7s $& 2.6309 & 0.881333 & 0.900734 & 2.17264 & 2.25313 & 3.02942 \\
106 &$ 5f^{14}  6d^4  7s  $&$ \;^{6}D_{1/2} $& 0.5376 & 0.8127 &$ 7s $& 2.5150 & 0.850008 & 0.867203 & 2.11798 & 2.21807 & 2.87277 \\
107 &$ 5f^{14}  6d^4  7s^2 $&$ \;^{5}D_{0} $& 0.5409 & 0.8240 &$ 7s $& 2.4351 & 0.820103 & 0.834045 & 1.95902 & 2.01018 & 2.80344 \\
108 &$ 5f^{14}  6d^5  7s^2 $&$ \;^{6}S_{5/2} $& 0.5396 & 0.8183 &$ 7s $& 2.3426 & 0.790744 & 0.807474 & 1.86004 & 1.95516 & 2.69468 \\
\hline
\hline
\end{tabular}
}
\end{center}
\end{table*}


\begin{table*}
\begin{center}
\caption{Correlation effects on the mean radius and mean spherical radius of No.\label{tab:rad-corr} }
\begin{tabular}{lrrrr}
\hline
\hline
	&	$<r>$ (a.u.)	&	var.	& $\sqrt{<r^2>}$ (a.u.)		& var.	\\
\hline
Ra$5f^{14} 7s^2 \;^1S_0$	&	0.574391	&		&	0.932605	&	\\
$+$ excit. of $7s \to 7p$, $6d$	&	0.573724	&	-0.12\%	&	0.928191	&	-0.47\% \\
$+$ excit. of $7s$, $5f \to 7p$, $6d$	&	0.573309	&	-0.19\%	&	0.925313	&	-0.78\%   \\
\hline
\hline
\end{tabular}
\end{center}
\end{table*}


\begin{table}
\begin{center}
\caption{Comparison between different definition of the ionic radii and experiment. We define the average radius of $5f$ orbitals as
$\bar{<r_{5f}}>=(6<r_{5f_{5/2}}>+8<r_{5f_{7/2}}>)/14$}
\label{tab:rad-exp}
\begin{tabular}{lrlrr}
\hline
\hline
	&	Exp.	&		&		&	Theo.	\\
\hline
$\Delta r_{\mathrm{Am}^+-\mathrm{Pu}^+} $\cite{bdhk2005} & -3.1\%  &($±$ 1.3\%) & $<r_{\mathrm{At.}}>$ & -2.1\% \\
            &    & &$\sqrt{<r^2_{\mathrm{At.}}>}$ & -3.1\% \\
            &    & &$\bar{<r_{5f}}>$ & -0.8\% \\
            &    & &$<r_{7s}>$ & 2.8\% \\
$\Delta r_{\mathrm{Fm}^+-\mathrm{Cf}^+}$ \cite{sbdk2003}& -2.0\% &  & $<r_{\mathrm{At.}}>$ & -1.7\% \\
            &    & &$\sqrt{<r^2_{\mathrm{At.}}>}$ & -2.4\% \\
            &    & &$\bar{<r_{5f}}>$ & -5.9\% \\
            &    & &$<r_{7s}>$ & -1.3\% \\
\hline
\hline
\end{tabular}
\end{center}
\end{table}

QED corrections on Landé factors of one and three electron atoms have been studied in great details
in the last few years, because of the increase in experimental accuracy associated with the
use of Penning traps in a new series of experiments \cite{hbhk2000,vdsv2004}. These experiments 
provided a strong incentives to evaluate very accurately QED corrections on Landé factors, beyond
the one due to the anomalous magnetic moment of the electron 
\cite{pssl1997,blps2000,yis2002,bisy2003,gstv2004,pjy2004,yis2004,pcjy2005}. To our knowledge, there has been only
one calculation dealing with QED corrections in heavy elements, in which the Feynman diagrams corresponding
to self-energy and other corrections were evaluated. This calculation concerned alkali elements up
 to francium ($Z=87$) \cite{lgp1999}.
 Here we deal with much more complex system, with several open shells,
and we calculate QED corrections due to the inclusion of the Breit interaction and of the vacuum polarization
in the SCF, as well as the contribution from the electron anomalous moment.
The coupling of an atom with a magnetic moment $\mu$ to a homogeneous magnetic field $\boldsymbol{B}$ 
 gives an energy change
\begin{equation}
\Delta  E = -\mu\cdot \boldsymbol{B}
\end{equation}
where $\mu=-g_{J}\mu_{B}\boldsymbol{J}$, $g_J$ being the Landé factor and $\mu_{B}$ the Bohr magneton.
The anomalous electron magnetic moment  corrections can be written (see, e.g., \cite{cac1985} and reference therein) as
\begin{equation}
\Delta g_{J} =(g-2) \frac{\langle J|| \beta \boldsymbol{\Sigma}||J\rangle }{\sqrt{J (J+1)(2J+1)}}\label{eq:gfacq}
\end{equation}
where $g=2\left(1+\frac{\alpha}{\pi}+\cdots \right)$ is the electron magnetic moment, $\beta$ and $\Sigma$ are $4× 4$ matrices, 
$\beta=\left(
\begin{array}{cc}
I & 0 \\
0 & I
\end{array}
\right)$
and
$\boldsymbol{\Sigma} =\left(
\begin{array}{cc}
\boldsymbol{\sigma}  & 0 \\
0 & \boldsymbol{\sigma} 
\end{array}
\right)$
where $I$ is a $2× 2$ identity matrix and $\boldsymbol{\sigma}$ are Pauli matrices.

As an example, we illustrate the different effects by evaluating the Landé $g$-factors for the ground configuration
of neutral actinide and transactinide up to $Z=106$, and of singly ionized lanthanide, actinide and
 transactinide up to $Z=106$. The results are displayed on Table \ref{table:gfac-neut} for neutral atoms and
 on Table \ref{table:gfac-sing-ion}, for singly ionized atoms. The ground configuration of neutral atom comes from \cite{dmcm2003}.
For singly ionized atoms, it has been taken from Refs.\ ~\cite{risp2004} for $Z$ up to 106 and from \cite{jfjd2002} for $Z=107$ and 108.
 Depending on the outer shell structures, 
the QED corrections can be dominated by either the $g-2$ correction or the Breit correction (order of 0.1\% of the
Landé factor). The vacuum polarization correction is at best one order of magnitude smaller at $Z=106$.

In Table \ref{table:gfac-exc-nob}, we present correlation effects on the Landé $g$-factor of the lowest levels of No, evaluated 
with the same wavefunctions as in Sec.\ ~\ref{sec:nobelium}, with the Breit operator in the SCF process.
 The effect of correlation ranges from 0.6\% to 0.02\% depending on the level.


\begin{table*}
\begin{center}
\caption{QED [Eq.\ \protect\eqref{eq:gfacq}], Breit and Uehling  corrections on Landé $g$-factors of neutral atoms with $90\leq Z\leq 106$ \label{table:gfac-neut} }
%
\begin{tabular}{rllrrrrrr}
\hline
\hline
Z &\multicolumn{1}{c}{ Conf.} & \multicolumn{1}{c}{Label} & \multicolumn{1}{c}{Landé (Coul.)} & \multicolumn{1}{c}{QED corr.} &
\multicolumn{1}{c}{ Breit contr.} &\multicolumn{1}{c}{ Uehl. corr.} & \multicolumn{1}{c}{Total }\\
\hline            
90 &$ 6d^2 7s^2         $&$ \;^3F_{2} $& 0.68158912 & -0.00073903 & -0.00031059 & -0.00002136 & 0.68051814 \\
91 &$ 5f^2 6d 7s^2          $&$ \;^4K_{11/2} $& 0.81847964 & -0.00042068 & -0.00020095 & -0.00001531 & 0.81784271 \\
92 &$ 5f^{3} 6d 7s^2          $&$ \;^5L_{6} $& 0.73962784 & -0.00060501 & -0.00090733 & -0.00001069 & 0.73810480 \\
93 &$ 5f^{4} 6d 7s^2          $&$ \;^6L_{11/2} $& 0.63735126 & -0.00084251 & -0.00108525 & -0.00000780 & 0.63541570 \\
94 &$ 5f^{6} 7s^2          $&$ \;^7F_{0} $&  &  &  &  \\
95 &$ 5f^{7} 7s^2         $&$ \;^8S_{7/2} $& 1.96702591 & 0.00224893 & 0.00193804 & 0.00000967 & 1.97122254 \\
96 &$ 5f^{7} 6d 7s^2         $&$ \;^9D_{2} $& 2.60336698 & 0.00372915 & 0.00336317 & 0.00002669 & 2.61048600 \\
97 &$ 5f^{9} 7s^2         $&$ \;^6H_{15/2} $& 1.29988703 & 0.00070147 & 0.00177562 & 0.00000642 & 1.30237054 \\
98 &$ 5f^{10} 7s^2          $&$ \;^5I_{8} $& 1.22357731 & 0.00052300 & 0.00108516 & 0.00000375 & 1.22518921 \\
99 &$ 5f^{11} 7s^2          $&$ \;^4I_{15/2} $& 1.18834647 & 0.00043997 & 0.00047015 & 0.00000159 & 1.18925817 \\
100 &$ 5f^{12} 7s^2          $&$ \;^3H_{6} $& 1.16189476 & 0.00037811 & 0.00019838 & 0.00000062 & 1.16247187 \\
101 &$ 5f^{13} 7s^2          $&$ \;^2F_{7/2} $& 1.14180915 & 0.00033132 & 0.00005605 & 0.00000020 & 1.14219672 \\
102 &$ 5f^{14} 7s^2          $&$ \;^1S_{0} $&  &  &  &  \\
103 &$ 5f^{14} 7s^2 7p $&$ \;^2P_{1/2} $& 0.66661695 & -0.00077309 & 0.00003007 & 0.00000003 & 0.66587396 \\
104 &$ 5f^{14} 6d^{2} 7s^2          $&$ \;^3P_{0} $&  &  &  &  \\
104 &$ 5f^{14} 7s^2 7p^2         $&$ \;^3F_{2} $& 0.69055158 & -0.00071837 & -0.00035975 & -0.00006392 & 0.68940954 \\
105 &$ 5f^{14}  6d^3 7s^2         $&$ \;^4F_{3/2} $& 0.45467514 & -0.00126763 & -0.00122595 & -0.00016406 & 0.45201750 \\
106 &$ 5f^{14} 6d^4 7s^2         $&$ \;^5D_{0} $&  &  &  &  \\
\hline
\hline
\end{tabular}
\end{center}
\end{table*}


\begin{table*}
\begin{center}
\caption{QED [Eq.\ \protect\eqref{eq:gfacq}], Breit and Uehling corrections on Landé $g$-factors of the ground configuration of 
singly ionized atoms with $57\leq Z\leq71$ and $90\leq Z\leq 108$.
 \label{table:gfac-sing-ion} }
%
\begin{tabular}{rllrrrrrr}
\hline
\hline
Z &\multicolumn{1}{c}{ Conf.} & \multicolumn{1}{c}{Label} & \multicolumn{1}{c}{Landé (Coul.)} & \multicolumn{1}{c}{QED corr.} &
\multicolumn{1}{c}{ Breit contr.} &\multicolumn{1}{c}{ Uehl. corr.} & \multicolumn{1}{c}{Total }\\
\hline            
57 &$ 5d^{2}        $&$ \;^{3}F_{2} $& 0.66973929 & -0.00076609 & -0.00012975 & -0.00000075 & 0.66884270 \\
58 &$ 4f 5d^{2}     $&$ \;^{4}H_{7/2} $& 0.71736392 & -0.00065375 & 0.00039406 & -0.00000787 & 0.71709635 \\
59 &$ 4f^{3} 6s          $&$ \;^{5}I_{4} $& 0.60262203 & -0.00092068 & -0.00031540 & -0.00000026 & 0.60138569 \\
60 &$ 4f^{4} 6s          $&$ \;^{6}I_{7/2} $& 0.44690832 & -0.00128150 & -0.00029522 & -0.00000023 & 0.44533137 \\
61 &$ 4f^{5} 6s          $&$ \;^{7}H_{2} $& 0.00677005 & -0.00230319 & -0.00086739 & -0.00000061 & 0.00359886 \\
62 &$ 4f^{6} 6s          $&$ \;^{8}F_{1/2} $& 3.95749786 & 0.00687585 & 0.00701619 & 0.00001509 & 3.97140499 \\
63 &$ 4f^{7} 6s          $&$ \;^{9}S_{4} $& 1.99480186 & 0.00230987 & 0.00035927 & 0.00000034 & 1.99747133 \\
64 &$ 4f^{7} 5d 6s          $&$ \;^{10}D_{5/2} $& 2.56089325 & 0.00362452 & 0.00067458 & 0.00000119 & 2.56519354 \\
65 &$ 4f^{9} 6s          $&$ \;^{7}H_{8} $& 1.36838651 & 0.00085813 & 0.00067487 & 0.00000040 & 1.36991992 \\
66 &$ 4f^{10} 6s          $&$ \;^{6}I_{17/2} $& 1.28727646 & 0.00067021 & 0.00068999 & 0.00000039 & 1.28863705 \\
67 &$ 4f^{11} 6s          $&$ \;^{5}I_{8} $& 1.24630282 & 0.00057447 & 0.00033068 & 0.00000018 & 1.24720815 \\
68 &$ 4f^{12} 6s          $&$ \;^{4}H_{13/2} $& 1.22865909 & 0.00053325 & 0.00015805 & 0.00000008 & 1.22935048 \\
69 &$ 4f^{13} 6s          $&$ \;^{3}F_{4} $& 1.24882255 & 0.00057982 & 0.00006252 & 0.00000004 & 1.24946493 \\
70 &$ 4f^{14} 6s          $&$ \;^{2}S_{1/2} $& 1.99992029 & 0.00231927 & -0.00000949 & -0.00000006 & 2.00223002 \\
71 &$ 4f^{14} 6d^{2}     $&$ \;^{1}S_{0} $&  &  &  &  \\
\hline               
90 &$ 5f^{2} 6d          $&$ \;^{4}K_{11/2} $& 0.80868853 & -0.00044409 & -0.00040145 & -0.00001738 & 0.80782561 \\
91 &$ 5f^{2} 6d 7s          $&$ \;^{5}K_{5} $& 0.69942415 & -0.00069727 & -0.00041498 & -0.00003386 & 0.69827803 \\
92 &$ 5f^{3} 6d 7s          $&$ \;^{6}L_{11/2} $& 0.63834785 & -0.00083965 & -0.00081792 & -0.00001647 & 0.63667382 \\
93 &$ 5f^{4} 6d 7s          $&$ \;^{7}L_{5} $& 0.52126261 & -0.00111157 & -0.00103625 & -0.00001135 & 0.51910344 \\
94 &$ 5f^{5} 6d 7s         $&$ \;^{8}K_{7/2} $& 0.26613829 & -0.00170714 & -0.00271582 & -0.00002210 & 0.26169324 \\
95 &$ 5f^{7} 7s       $&$ \;^{9}S_{4} $& 1.97238902 & 0.00226041 & 0.00159104 & 0.00001036 & 1.97625083 \\
96 &$ 5f^{7} 7s^{2}         $&$ \;^{8}S_{7/2} $& 1.95878066 & 0.00223118 & 0.00242861 & 0.00001187 & 1.96345232 \\
97 &$ 5f^{8} 6d 7s         $&$ \;^{9}G_{7} $& 1.48746928 & 0.00113601 & 0.00161739 & 0.00001838 & 1.49024105 \\
98 &$ 5f^{10} 7s          $&$ \;^{6}I_{17/2} $& 1.27036131 & 0.00063117 & 0.00097651 & 0.00000581 & 1.27197480 \\
99 &$ 5f^{11} 7s          $&$ \;^{5}I_{8} $& 1.23940637 & 0.00055819 & 0.00042388 & 0.00000222 & 1.24039065 \\
100 &$ 5f^{12} 7s          $&$ \;^{2}H_{11/2} $& 1.09903182 & 0.00023261 & 0.00025286 & 0.00000531 & 1.09952260 \\
101 &$ 5f^{13} 7s          $&$ \;^{3}F_{4} $& 1.24906758 & 0.00057982 & 0.00004304 & 0.00000013 & 1.24969057 \\
102 &$ 5f^{14} 7s          $&$ \;^{2}S_{1/2} $& 1.99988521 & 0.00231927 & -0.00002001 & -0.00000035 & 2.00218413 \\
103 &$ 5f^{14} 7s^2          $&$ \;^{1}S_{0} $&  &  &  &  \\
104 &$ 5f^{14} 6d 7s^2          $&$ \;^{2}D_{3/2} $& 0.79977826 & -0.00046385 & 0.00005358 & 0.00000012 & 0.79936810 \\
105 &$ 5f^{14}  6d^2   7s^2      $&$ \;^{3}F_{2} $& 0.70320927 & -0.00068901 & -0.00043211 & -0.00005473 & 0.70203342 \\
106 &$ 5f^{14}  6d^4  7s  $&$ \;^{6}D_{1/2} $& 3.19780436 & 0.00510655 & 0.00311970 & 0.00051784 & 3.20654845 \\
107 &$ 5f^{14}  6d^4  7s^2 $&$ \;^{5}D_{0} $&  &  &  &  \\
108 &$ 5f^{14}  6d^5  7s^2 $&$ \;^{6}S_{5/2} $& 1.87088576 & 0.00203080 & 0.00399742 & 0.00034326 & 1.87725723 \\
\hline
\hline
\end{tabular}
\end{center}
\end{table*}

\begin{table}
\begin{center}
\caption{Correlation effect on the Landé $g$-factor of the first excited states of nobelium.
 \label{table:gfac-exc-nob} }
%
\begin{tabular}{lrrr}
\hline
\hline
Level & \multicolumn{1}{c}{$7s7p \;^3P_1$} & \multicolumn{1}{c}{$7s7p \;^1P_1$} & \multicolumn{1}{c}{$7s7p \;^3P_2$} \\
\hline
DF & 1.49073 & 1.00932 & 1.50111 \\
$7$, $7p$ exc. $\to 7p$, $6d$ & 1.47691 & 1.02829 & 1.50027 \\
$7$, $7p$, $5f$ exc. $\to 7p$, $6d$ & 1.48845 & 1.01507 & 1.50080 \\
\hline
\hline
\end{tabular}
\end{center}
\end{table}

\section{Conclusions}
\label{sec:concl}

In this work we have evaluated the effects of self-consistent Breit interaction and vacuum polarization on level energies, transition
energies and probabilities Landé $g$-factors on super-heavy elements. We have also studied their effect on orbitals and atomic charge distribution radii and
 found very large effects on highly charged ions. We have found some hints, when studying Be-like ions, of rather strong non-perturbative correlation effects
for $Z\approx 128$. For neutral or quasi-neutral systems, self-consistent Breit interaction and vacuum polarization have a small but noticeable effects on Landé factor and transition rates, which could be
felt experimentally. Transition energies, on the other hand, are heavily dominated by Coulomb correlation. This is rather good news, since treating the Breit
interaction self-consistently obliges to evaluate magnetic and retardation integrals during the SCF process, which are about one order
of magnitude more numerous than Coulomb ones, leading to calculations that cannot fit on even on the largest computers available today,
 even with relatively small configuration space.

 We have shown that very large non-relativistic offset may affect the fine structure separation of elements with several open shells.
This should be carefully taken into account to avoid providing completely wrong results with  all-order methods. We have also shown that the inclusion of the
Breit interaction in the SCF process, because it get mixing coefficients closer to the jj limit, can noticeably complicate numerical convergence.

\section*{Acknowledgments}
\label{sec:5}

Laboratoire Kastler Brossel is Unit{é} Mixte de Recherche du CNRS
n$^{\circ}$ C8552. This research was partially supported by the FCT projects
POCTI/FAT/44279/2002 and POCTI­/0303/2003  (Portugal),
financed by the European Community Fund FEDER, and by the French-Portuguese collaboration (PESSOA Program, Contract n$^{\circ}$ 10721NF).
One of us (P.I.) thanks M. Sewtz for several discussions on this subject.
The 8-processors workstation used for the calculations has been provided by
 an ``infrastructure'' grant from the ``Ministère de La Recherche et de l'Enseignement Supérieur''.
%

\bibliography{refs}

\end{document}